\documentclass[iop]{emulateapj}

\usepackage{amsmath,natbib,graphicx}
\usepackage{epsf}
\usepackage{epstopdf}
\usepackage{epsfig}
\usepackage{color}
\DeclareGraphicsExtensions{.jpg,.pdf,.png,.eps,.ps}
\usepackage{scrextend}
\usepackage{hyperref}

\def\UFlorida{1}
\def\CfA{4}
\def\Illinois{15}
\def\Arizona{2}
\def\MPIfR{3}
\def\ESOGarching{6}
\def\Stanford{12}
\def\UCLA{13}
\def\Dal{9}
\def\Diego{5}
\def\Cambridge{7}
\def\Davis{10}
\def\UCL{11}
\def\Case{14}
\def\Miss{8}
\def\Oxford{16}

\begin{document}

\title{Stellar masses and star formation rates of lensed dusty star-forming galaxies from the SPT survey}



\shortauthors{J. Ma, et al.}

\author{
Jingzhe Ma$^{\UFlorida}$,
Anthony.~H.~Gonzalez$^{\UFlorida}$, 
J.~S.~Spilker$^{\Arizona}$,
M.~Strandet$^{\MPIfR}$, 
M.~L.~N.~Ashby$^{\CfA}$,
M.~Aravena$^{\Diego}$,
M.~B\'ethermin$^{\ESOGarching}$,
M.~S.~Bothwell$^{\Cambridge}$,
C.~de~Breuck$^{\ESOGarching}$,
M.~Brodwin$^{\Miss}$,
S.~C.~Chapman$^{\Dal}$,
C.~D.~Fassnacht$^{\Davis}$,
T.~R.~Greve$^{\UCL}$,	
B.~Gullberg$^{\ESOGarching}$, 
Y.~Hezaveh$^{\Stanford}$,
M.~Malkan$^{\UCLA}$,
D.~P.~Marrone$^{\Arizona}$,  
B.~R.~Saliwanchik$^{\Case}$,
J.~D.~Vieira$^{\Illinois}$,
A.~Wei\ss$^{\MPIfR}$,
N.~Welikala$^{\Oxford}$
}

\altaffiltext{\UFlorida}{Department of Astronomy, University of Florida, Gainesville, FL 32611, USA; \href{mailto:jingzhema@ufl.edu}{jingzhema@ufl.edu}}
\altaffiltext{\Arizona}{Steward Observatory, University of Arizona, 933 North Cherry Avenue, Tucson, AZ 85721, USA}
\altaffiltext{\MPIfR}{Max-Planck-Institut f\"{u}r Radioastronomie, Auf dem H\"{u}gel 69 D-53121 Bonn, Germany}
\altaffiltext{\CfA}{Harvard-Smithsonian Center for Astrophysics, 60 Garden Street, Cambridge, MA 02138, USA}
\altaffiltext{\Diego}{N\'ucleo de Astronom\'{\i}a, Facultad de Ingenier\'{\i}a, Universidad Diego Portales, Av. Ej\'ercito 441, Santiago, Chile}
\altaffiltext{\ESOGarching}{European Southern Observatory, Karl Schwarzschild Stra\ss e 2, 85748 Garching, Germany}
\altaffiltext{\Cambridge}{Cavendish Laboratory, University of Cambridge, JJ Thompson Ave, Cambridge CB3 0HA, UK}
\altaffiltext{\Dal}{Dalhousie University, Halifax, Nova Scotia, Canada}
\altaffiltext{\Davis}{Department of Physics,  University of California, One Shields Avenue, Davis, CA 95616, USA}
\altaffiltext{\UCL}{Department of Physics and Astronomy, University College London, Gower Street, London WC1E 6BT, UK}
\altaffiltext{\Stanford}{Kavli Institute for Particle Astrophysics and Cosmology, Stanford University, Stanford, CA 94305, USA}
\altaffiltext{\UCLA}{Department of Physics and Astronomy, University of California, Los Angeles, CA 90095-1547, USA}
\altaffiltext{\Illinois}{Department of Astronomy and Department of Physics, University of Illinois, 1002 West Green St., Urbana, IL 61801, USA}
\altaffiltext{\Case}{Department of Physics, Case Western Reserve University, Cleveland, Ohio 44106, USA}
\altaffiltext{\Miss}{Department of Physics and Astronomy, University of Missouri, 5110 Rockhill Road, Kansas City, MO 64110, USA}
\altaffiltext{\Oxford}{Department of Physics, Oxford University, Denis Wilkinson Building, Keble Road, Oxford, OX1 3RH, UK}


\begin{abstract}
To understand cosmic mass assembly in the Universe at early epochs, we primarily rely on measurements of stellar mass and star formation rate of distant galaxies. In this paper, we present stellar masses and star formation rates of six high-redshift ($2.8\leq z \leq 5.7$) dusty, star-forming galaxies (DSFGs) that are strongly gravitationally lensed  by foreground galaxies. These sources were first discovered by the South Pole Telescope (SPT) at millimeter wavelengths and all have spectroscopic redshifts and robust lens models derived from ALMA observations. We have conducted follow-up observations, obtaining multi-wavelength imaging data, using  {\it HST}, {\it Spitzer}, {\it Herschel} and the Atacama Pathfinder EXperiment (APEX). We use the high-resolution {\it HST}/WFC3 images to disentangle the background source from the foreground lens in {\it Spitzer}/IRAC data. The detections and upper limits provide important constraints on the spectral energy distributions (SEDs) for these DSFGs, yielding stellar masses, IR luminosities, and star formation rates (SFRs). The SED fits of six SPT sources show that the intrinsic stellar masses span a range more than one order of magnitude with a median value $\sim$ 5 $\times 10^{10}M_{\Sun}$. The intrinsic IR luminosities range from 4$\times 10^{12}L_{\Sun}$ to 4$\times 10^{13}L_{\Sun}$. They all have prodigious intrinsic star formation rates of 510 to 4800 $M_{\Sun} {\rm yr}^{-1}$.  Compared to the star-forming main sequence (MS), these six DSFGs have specific SFRs that all lie above the MS, including two galaxies that are a factor of 10 higher than the MS. Our results suggest that we are witnessing the ongoing strong starburst events which may be driven by major mergers.

\end{abstract}
\section{Introduction}\label{sec:intro}

Studies of the Cosmic Infrared Background have shown that about half of the energy produced in cosmic history comes from distant dust-enshrouded star-forming galaxies \citep{dole06}.  Ever since the first extragalactic submillimeter-wavelength surveys \citep{smail97,hughes98,barger98} were carried out with the Submillimeter Common User Bolometer Array (SCUBA; \citealt{holland99}) at 850$\mu$m, the existence of a population of high-redshift dusty star-forming galaxies (DSFGs) has been established, expanding our understanding of galaxy formation and evolution. These massive galaxies ($M_{*}\sim10^{11}M_{\Sun}$; \citealt{hainline11,michalowski12,simpson14}) are extremely bright at submillimeter wavelengths but faint in the optical due to extinction and hence are also known as ``submillimeter galaxies" (SMGs).  Their spectral energy distributions (SEDs) are dominated by rest-frame far-infrared luminosities in excess of $10^{12}L_{\Sun}$ (e.g., \citealt{coppin08,Magnelli12}). These dusty luminous galaxies are producing stars at typically intense star formation rates of $>100-1000$ $M_{\Sun}{\rm yr}^{-1}$ (e.g., \citealt{neri03,chapman05}). The high SFRs can be explained by either large reservoirs of molecular gas or a boosted star formation efficiency induced by major mergers \citep{narayanan09,engel10,fu13,Bethermin15a,Dye15}. The dominant mechanism is still a topic of debate.

Previous studies of their stellar masses and prodigious star formation rates have been limited to the most luminous systems due to the challenge of obtaining redshifts and detailed follow-up observations of these distant, dust-obscured objects.  The South Pole Telescope (SPT; \citealt{carlstrom11}), which surveyed  2500 ${\rm deg}^2$ at 1.4, 2, and 3mm with high spatial resolution (FWHM $\sim$1.0$\arcmin$), has helped us overcome these difficulties by discovering $\sim$100 high redshift, strongly gravitationally lensed DSFGs (\citealt{vieira10,mocanu13}).   The lensing origin of the SPT DSFGs was confirmed by the Atacama Large Millimeter/submillimeter Array (ALMA) (\citealt{hezaveh13,vieira13}). The ALMA submillimeter imaging is used to model the lensing geometry yielding magnification factors up to 30 (\citealt{hezaveh13}; Spilker et al. in prep). Strong gravitational lensing enables detailed analyses of these DSFGs but also complicates recovery of the properties of the sources because emission from the background DSFG is blended and confused with that from the foreground lensing galaxy. In the FIR/submillimeter regime, the background emission dominates, but in the rest-frame optical/near-IR where we probe the established stellar population, we must de-blend the background DSFG from the foreground lens. 

The primary goal of this paper is to measure stellar masses and hence determine the specific SFRs (sSFRs), which are a critical diagnostic for understanding the nature of these galaxies. We develop and implement an image de-blending technique with which we can recover the stellar mass of the lensed DSFG. With the lens model, we are able to derive the intrinsic properties of the SPT DSFGs (e.g., stellar mass, luminosity, and star formation rate) and place these galaxies in the context of the evolution of the high-redshift galaxy population as a whole. We present six SPT DSFGs all of which have been systematically followed up with {\it HST}, {\it Spitzer}, APEX, and  {\it Herschel} and have robust lens models derived from high resolution ALMA imaging.

This paper is organized as follows. In Section \ref{sec:observations}, we present our multi-wavelength observations.  Section \ref{sec:deblending} describes the de-blending technique and results for each source. In Section \ref{sec:lensmodel}, we describe the lens model.  In Section \ref{sec:SEDfitting}, we present the stellar masses and star formation rates derived from SED fitting.  We discuss the SFR -- M$_*$ relation and sSFRs in Section \ref{sec:discussion}. A summary of the feasibility analysis of the de-blending technique and conclusions are presented in Section \ref{sec:conclusions}. Throughout this work, we use a standard cosmological model with $H_0=70$ ${\rm km}$ ${\rm s}^{-1}$${\rm Mpc}^{-1}$, $\Omega_M=0.3$, and $\Omega_{\Lambda}=0.7$.

\section{Observations}
\label{sec:observations}
\begin{figure*}
\centering
{\includegraphics[width=12cm, height=8cm]{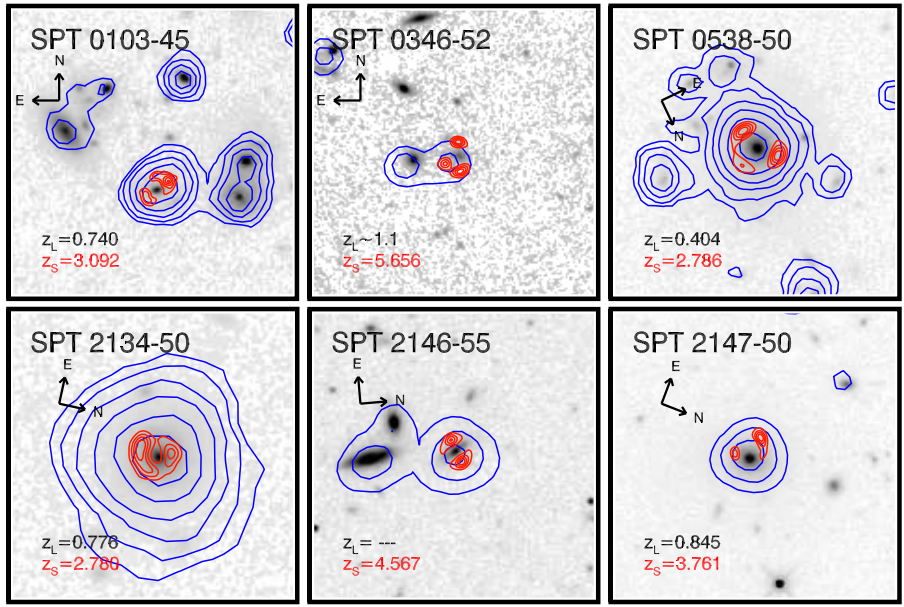}} 
\caption{20$\arcsec$ $\times$ 20$\arcsec$ cutouts of the six SPT DSFGs showing the {\it HST}/WFC3 (grey), {\it Spitzer}/IRAC (blue contours), and ALMA Band 7 (red contours) data. }
\label{fig:fig1}
\end{figure*}

The six SPT DSFGs presented in this paper (Tables \ref{tab:obs} and \ref{tab:params}) are drawn from a parent sample of 50 SPT DSFGs with {\it Spitzer} imaging, and constitute the best cases among the subset (13 sources ) with {\it HST} data and lens models (Sec. \ref{sec:lensmodel}) derived from ALMA imaging. Fig. \ref{fig:fig1} shows the {\it HST}, {\it Spitzer}, and ALMA data for the six sources. 

\subsection{Spitzer/IRAC}

A total of 50 SPT DSFGs were followed up by two Spitzer Space Telescope programs (PID 60194 and PID 80221; PI Vieira) and a joint {\it Spitzer}/{\it HST} program (PID 10094; PI Vieira) using the Infrared Array Camera (IRAC; \citealt{fazio04}). For the Spitzer programs, images were taken at 3.6$\mu$m and 4.5$\mu$m with 36 or 32 (dither pattern) $\times$ 100 second (frame time)  exposures and 12 (dither pattern) $\times$30  (frame time)  exposures, respectively. For the joint program, we took the 100s frames in both bands with 36/72/108 dither positions. We summarize the data for each source in Table \ref{tab:obs}. The basic calibrated data, pre-processed using the standard pipeline by the Spitzer Science Center, were resampled and combined into a mosaic image utilizing the MOPEX software package \citep{makovoz05} and IRACproc \citep{schuster06}. The IRAC mosaics have a resampled pixel scale of  $0.6''$/pixel and an angular resolution of $\sim 1.7''$.  The IRAC data are the primary probes of the stellar component of the background DSFGs. However, due to the close proximity of the foreground lensing galaxy, emission from the lensed DSFG is mixed with that of the foreground lens and needs to be de-blended. The de-blending technique, which will be described in Sec.\ref{sec:deblending},  is  a critical component of this project, making extracting fluxes from extremely distant sub-millimeter sources possible.

\begin{table*}
\centering
\caption{Summary of Spitzer/IRAC and HST/WFC3 data}
\begin{tabular}{@{}lclcc@{}}
\hline
\hline
SPT Source     &  Short Name &Filters (exposure time) & Spitzer PID & HST PID \\
\hline
SPT-S J010312-4538.8 & SPT0103-45     & F110W (1312s),  F160W (1412s), 3.6$\mu$m (3600s), 4.5$\mu$m (7200s)  &  10094       &  12659 \\
SPT-S J034640-5204.9 &	SPT0346-52	& F110W (1312s),  F160W (1412s), 3.6$\mu$m (3600s), 4.5$\mu$m (10800s) & 10094      & 12659 \\
SPT-S J053816-5030.8 &	SPT0538-50	& F110W (1312s),  F160W (1412s), 3.6$\mu$m (3600s), 4.5$\mu$m (360s)   &   60194     &  12659  \\
SPT-S J213404-5013.2 & SPT2134-50	& F110W (1312s),  F160W (1412s), 3.6$\mu$m (3200s), 4.5$\mu$m (360s)    &  80221       & 12659 \\
SPT-S J214654-5507.8 & SPT2146-55     & F140W (2812s), 3.6$\mu$m (3600s), 4.5$\mu$m (3600s)                             & 10094      &   13614\\  
SPT-S J214720-5035.9 & SPT2147-50     & F140W (2812s), 3.6$\mu$m (3600s), 4.5$\mu$m (3600s)                             & 10094      &   13614 \\   
\hline

\hline
\end{tabular}
\label{tab:obs}
\end{table*}

\subsection{{\it HST}/WFC3-IR}

A sample of 18 SPT sources were observed with the  Wide Field Camera 3 (WFC3) on board the Hubble Space Telescope under program 12659 (PI: Vieira).  Each target was assigned one complete orbit split between F110W and F160W filters in the infrared channel. Another six sources were followed up in the joint {\it Spitzer}/{\it HST} program (PID 13614; PI: Vieira) at F140W with one orbit assigned per source. Dithering was implemented for maximum resolution.  The data were reduced using the standard {\it HST} pipeline. The pixel size of the WFC3 images is 0.128$\arcsec$. Detailed information on the WFC3 data products can be found in the WFC3 instrument handbook \citep{Dressel15}.  

\subsection{ALMA imaging}

The gravitationally lensed origin for the SPT sources was confirmed by ALMA under a Cycle 0 program (2011.0.00958.S; PI: D. Marrone). We imaged 47 SPT sources with ALMA at 870$\mu$m with the dual-polarization Band 7 (275-373 GHz) receivers in both compact and extended array configurations,  reaching $\sim$ 0.5$\arcsec$ resolution.  We refer the reader to \cite{hezaveh13} for details of the ALMA observations. The visibilities from each configuration were concatenated and imaged using uniform weighting to emphasize resolved structure. The combined ALMA images are used in this project to indicate position and structure of the DSFGs. 

\subsection{FIR data}

\subsubsection{APEX/LABOCA}

The SPT sources were imaged at 870$\mu$m with the Large APEX BOlometer CAmera (LABOCA) at APEX under the program M-085.F-0008 (2010, PI:Weiss). LABOCA is a 295-element bolometer array (Siringo et al. 2009 ) with an 11.4$\arcmin$  field-of-view and an angular resolution of 19.7$\arcsec$ (FWHM). The central frequency of LABOCA is 345GHz (870$\mu$m) with a passband FWHM of $\sim$ 60GHz. Observations were carried out under good weather conditions (typical PWV of 0.9mm ranging between 0.3 and 1.5mm). The data reduction was performed in the same manner as in \cite{greve12}.

\subsubsection{Herschel/SPIRE}

The SPIRE \citep{griffin10} data were obtained in the programs $OT1\_jvieira\_4$ and $OT2\_jvieira\_5$, and were observed simultaneously at 250$\mu$m, 350$\mu$m and 500$\mu$m. The maps were produced using the standard reduction pipeline HIPE v9.0. The fluxes were extracted by fitting a Gaussian to the source and reading out the maximum value. The fluxes have been corrected for pixelization as described in the SPIRE Observers' Manual \citep{Valtchanov14}. 

\subsubsection{Herschel/PACS}

The PACS \citep{griffin10} data at 100$\mu$m and 160$\mu$m were obtained in the programs $OT1\_jvieira\_4$, $OT1\_dmarrone\_1$ and $OT2\_jvieira\_5$.  Each PACS map has been co-added, weighted by coverage. We performed aperture photometry with the aperture sizes fixed to 7$\arcsec$ for the 100$\mu$m map and 10$\arcsec$ for the 160$\mu$m map. Details of the FIR data observations, reduction, and photometry will be given in Strandet et al. (in preparation). 

\subsection{ALMA redshift survey}

Using ALMA, we conducted a blind redshift survey in the 3mm atmospheric transmission window for a subset of the SPT sources. The lines detected are identified as redshifted emission lines of $^{12}$CO, $^{13}$CO, CI, H$_2$O, and H$_2$O$^+$. The details of the spectral line observations and the redshift distribution of dusty star-forming galaxies by SPT can be found in \cite{weiss13}.  We present in Table \ref{tab:params} the redshifts  for the sources analyzed in this paper. 

\begin{table*}
\centering
\caption{Lens and source parameters}
\begin{tabular}{@{}lcccccl@{}}
\hline
\hline
SPT source name   & $z_L$ &  $z_S$  &  $\mu$                              & S\'ersic & ExpDisk ($n$ = 1) & PSF\\ 
                                  &            &            &                                          & ($n$, r$_e$, mag)  & (r$_e$, mag)  & (mag) \\                                                        
\hline                        												
SPT0103-45\footnote{\label{a}Magnitudes (AB) measured in the F110W filter.}          &  0.740      & 3.092            & 5.34 $\pm$ 0.11     &    (1.9,  0.3$\arcsec$, 21.84)  &---  &     23.89  \\                     
SPT0346-52\footref{a}   	      & $\sim$1.1	& 5.656            & 5.57 $\pm$ 0.12      &     (2.3, 0.6$\arcsec$, 24.58)   &--- &   ---     \\					
SPT0538-50\footref{a}    	      & 0.404	& 2.786    	        & 20.12 $\pm$ 1.81    &(3.2, 0.7$\arcsec$, 18.58)   &(2.4$\arcsec$, 19.74) &  20.97 \\					     
SPT2134-50\footref{a}  	      & 0.776	& 2.780            & 21.00 $\pm$ 2.42     &  (1.9, 0.6$\arcsec$, 20.17) &---  & 21.88 \\					
SPT2146-55\footnote{\label{b}Magnitudes (AB) measured in the F140W filter.}           & ---            & 4.567            & 6.65 $\pm$ 0.41       & (3.1, 0.1$\arcsec$, 23.77) &    (0.6$\arcsec$, 22.09)& ---\\					     
SPT2147-50\footref{b}          & 0.845       & 3.761            & 6.55 $\pm$ 0.42       & (2.7, 0.4$\arcsec$, 20.02) &  (1.3$\arcsec$, 20.80)&--- \\
\hline

\hline
\end{tabular}
\tablecomments{We adopt the source redshift for SPT0538-50 obtained with ATCA CO(1-0) spectroscopy by \cite{aravena13}. We use the lensing magnification factors in Spilker et al. (in preparation). The lens parameters (e.g., S\'ersic index, effective radius) derived from F110W or F140W images using GALFIT  are held fixed during the de-blending process. We list the parameters for each component: the S\'ersic profile, the ExpDisk (i.e., exponential disk), and the PSF point source component.} 
\label{tab:params}

\end{table*}

\section{De-blending technique}
\label{sec:deblending}

\subsection{Modeling the lenses in {\it HST}}

To construct models of the surface brightness profiles for each lensing galaxy, we perform profile fitting using GALFIT \citep{peng10}, which can simultaneously fit profiles to multiple galaxies. We start with a single S\'ersic profile \citep{Sersic63} and then add other components to generate better models as needed. We create the models using F110W filter images when available, as F110W contains less emission from the background DSFG  than F160W and therefore enables better modeling of the foreground lens.  \newline

Here we present the modeling results for the six systems in our DSFG sample.  SPT0538-50, which has been modeled in \cite{bothwell13b}, serves as a test case for our analysis. For the non-detections, we place upper limits on the flux densities to constrain the SED of the background source. To derive photometric upper limits, we performed random aperture photometry on the background using a radius for each source that encloses the 3$\sigma$ ALMA contours. The detections and $3\sigma$ upper limits are presented in Table \ref{tab:photometry}.\\

\subsubsection{SPT0538-50}

\begin{figure*}
\centering
{\includegraphics[width=17cm, height=15.5cm]{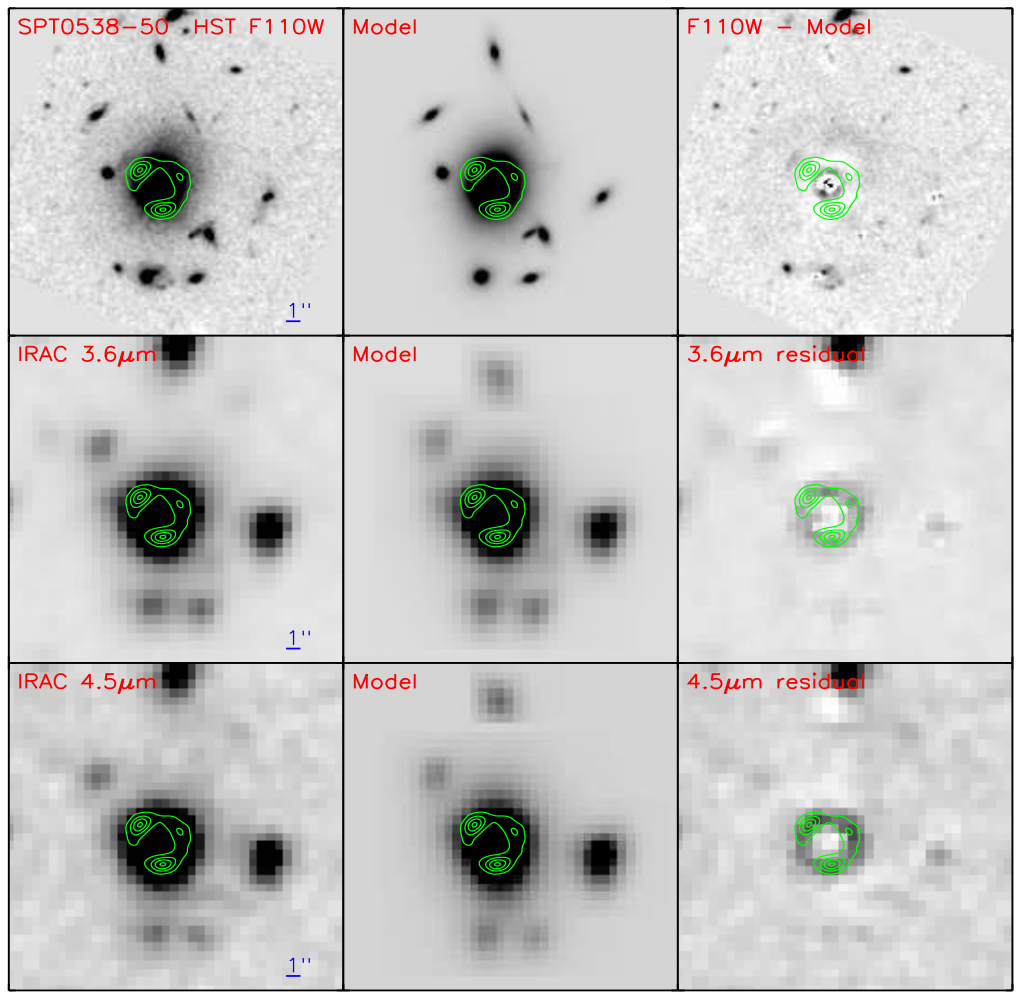}} 
\caption{23$\arcsec$ $\times$ 23$\arcsec$ {\it HST} and IRAC cutouts of SPT0538-50 showing the de-blending technique. Each image is oriented such that North is up and East is to the left.  In each panel, the green contours show ALMA sub-millimeter continuum emission, indicating the position and structure of the background dusty star-forming galaxy.  Contours are in steps of 5$\sigma$, starting at 3$\sigma$. The top row shows modeling of the lens in {\it HST} WFC3 F110W.  {\it Top Left:} The original {\it HST}/WFC3 image of SPT0538-50 in F110W filter. {\it Top Middle:}  The best-fit GALFIT model of SPT0538-50. {\it Top Right:} The residual image after subtracting the best-fit model from the original {\it HST} image. The residual flux of the knot south-east away from the center is 0.3\%.\\
The middle row and bottom row show de-blending of the source/lens in IRAC at 3.6$\mu$m and 4.5$\mu$m, respectively. {\it Left:}  The original Spitzer/IRAC 3.6$\mu$m/4.5$\mu$m image of SPT0538-50. {\it Middle:}  The GALFIT model convolved with the IRAC 3.6$\mu$m/4.5$\mu$m PSF.  {\it Right:} The residual image. The Einstein ring is recovered in IRAC after the foreground lens is subtracted off.  }
\label{fig:0538}
\end{figure*}

We first simultaneously fit the central lensing galaxy and 9 neighboring objects  with a single S\'ersic profile for each of them as a first pass. The resultant best-fit model for the central lens has a S\'ersic index (i.e., the shape parameter controlling the central concentration of the profile) of  $n$ = 8.8 and large residuals with significant positive and negative counts left over in the center. We improve the fit by using a three-component model for the central lensing galaxy, consisting of a S\'ersic profile, plus an exponential disk, plus a point source (SEP). A S\'ersic bulge and an exponential disk are insufficient and the point source is needed to minimize the residuals. The SEP model yields reasonable residuals with $n$ = 3.2. We present the fitting result of the SEP model in Fig. \ref{fig:0538}. The top panels of Fig. \ref{fig:0538} show the original WFC3 image (top left), the best-fit model by GALFIT (top middle), and the residuals leftover after subtracting the best-fit model from the original image (top right). We see no excess emission associated with the structure indicated by the ALMA contours, suggesting the fluxes extracted from the WFC3  F110W image are only from the foreground lensing galaxy. Adding additional components does not substantially improve the fit, and is not physically motivated. We therefore continue the de-blending process with the three-component SEP model. Note that in \cite{bothwell13b} the lensing galaxy was modeled using a co-add between the F110W and F160W bands while we use the bluest band throughout the paper. 

\subsubsection{SPT0103-45}

\begin{figure*}
\centering
{\includegraphics[width=17cm, height=15.5cm]{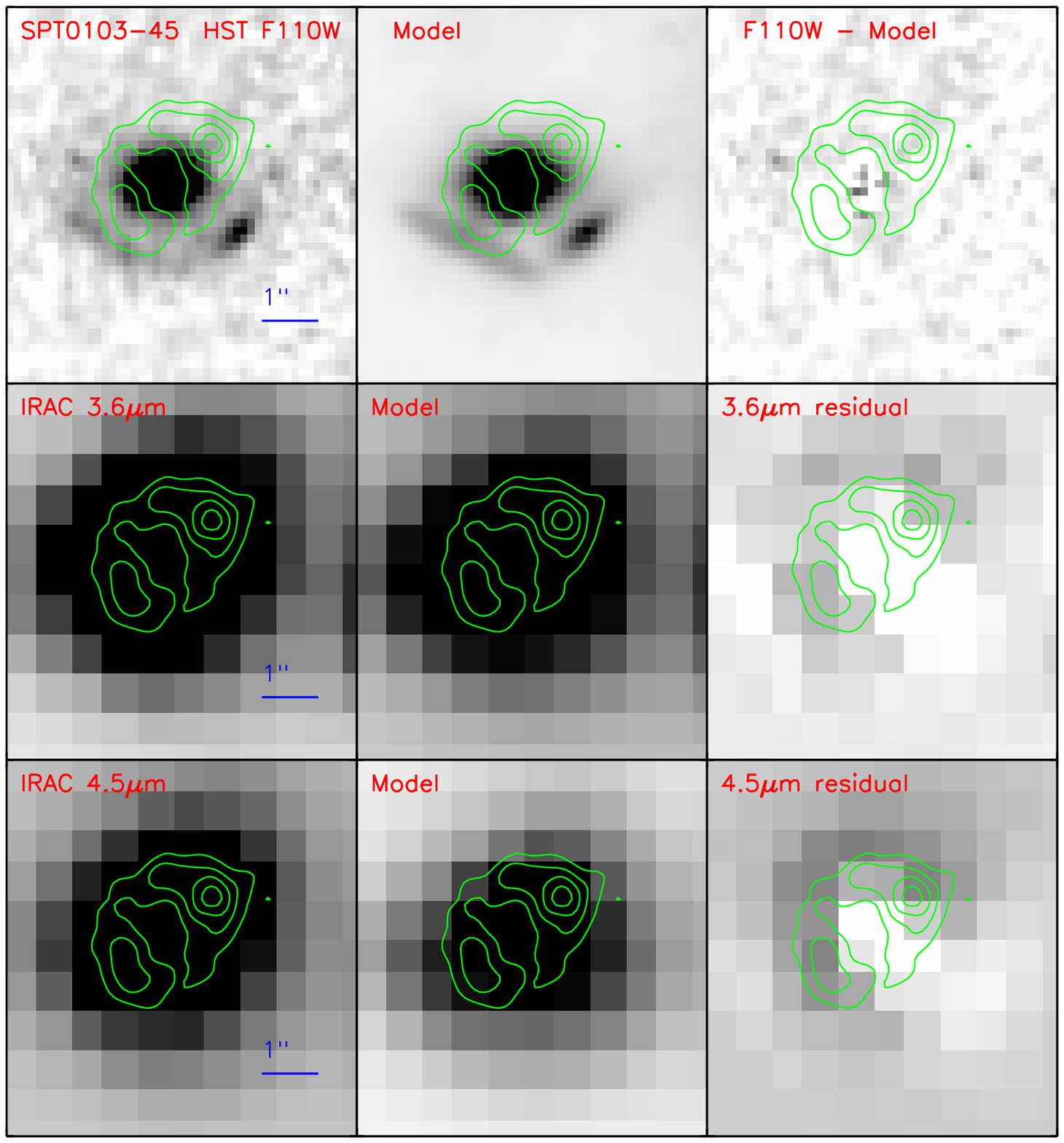}} 
\caption{6$\arcsec$ $\times$ 6$\arcsec$ {\it HST} and IRAC cutouts of SPT0103-45 showing the de-blending technique. Subfigures are the same as in Figure \ref{fig:0538}. }
\label{fig:0103}
\end{figure*}

The source is best fit with a S\'ersic model ($n$ = 1.9) and a point source. The obvious ``arc"-like structure (Fig. \ref{fig:0103} top left) is not the lensing arc associated with the background DSFG, which is indicated by the ALMA contours. We treat that as two elongated S\'ersic-profile galaxies that are subtracted off from the original image. The top middle panel shows the best-fit model by GALFIT and the overall residuals after the subtraction are shown in the top right panel.

\subsubsection{SPT0346-52}

\begin{figure*}
\centering
{\includegraphics[width=17cm, height=15.5cm]{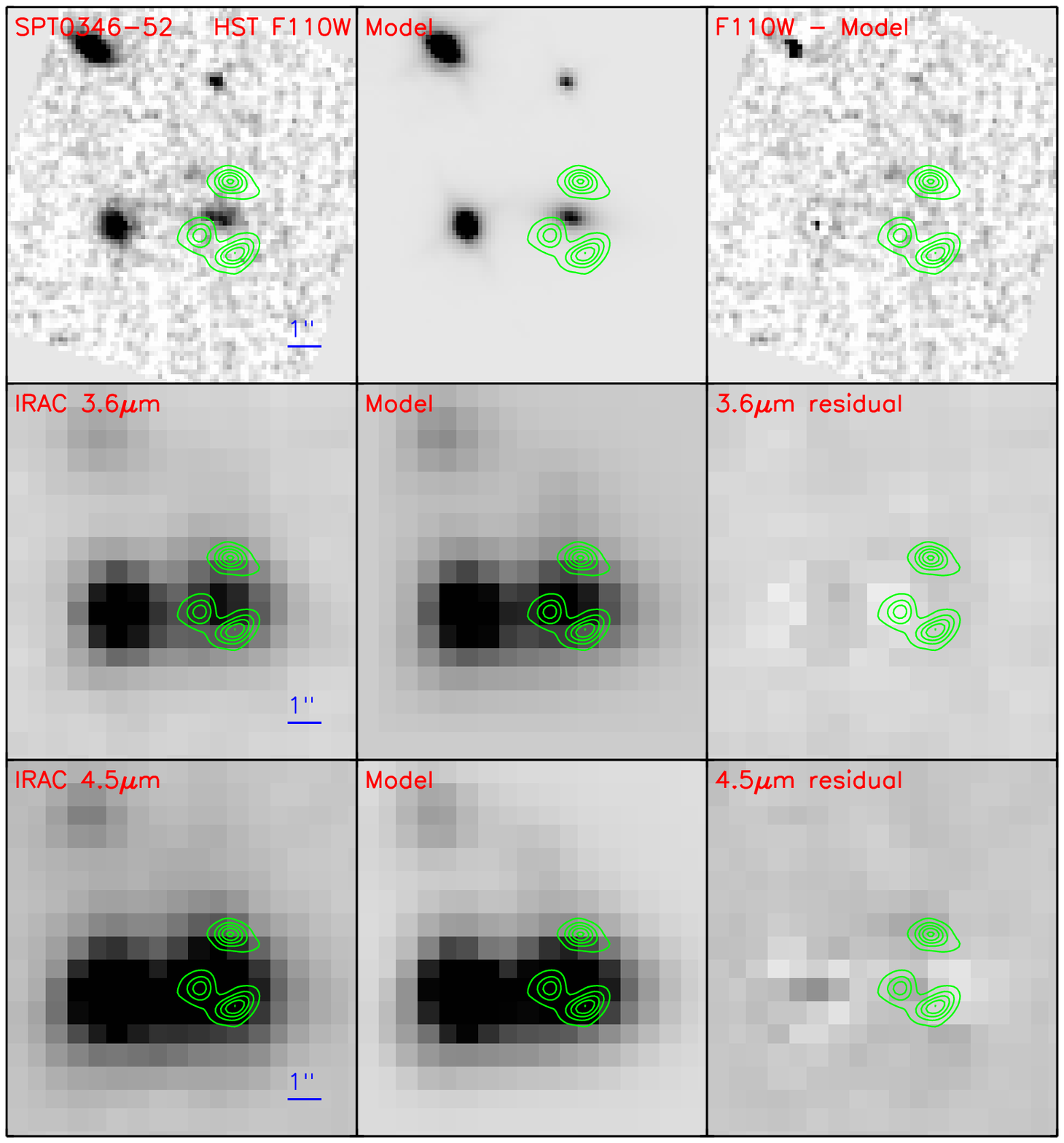}} 
\caption{10$\arcsec$ $\times$ 10$\arcsec$ {\it HST} and IRAC cutouts of SPT0346-52 showing the de-blending technique.  Subfigures are the same as in Figure \ref{fig:0538}. }
\label{fig:0346}
\end{figure*}

SPT0346-52 has the highest redshift ($z$ = 5.7) among all the sources in this work.  We fit four galaxies in the field with a single S\'ersic profile ($n$ = 2.3 for the lensing galaxy) by GALFIT, shown in Fig.\ref{fig:0346} (top panels), yielding a clean residual. Given its high redshift, it is not surprising that SPT0346-52 is undetected in the F110W band. We note that the arcs are well-separated from the central lens with a diameter of 2.2$\arcsec$, allowing a de-coupling of the background and foreground emission in IRAC. 

\subsubsection{SPT2134-50}

\begin{figure*}
\centering
{\includegraphics[width=17cm, height=15.5cm]{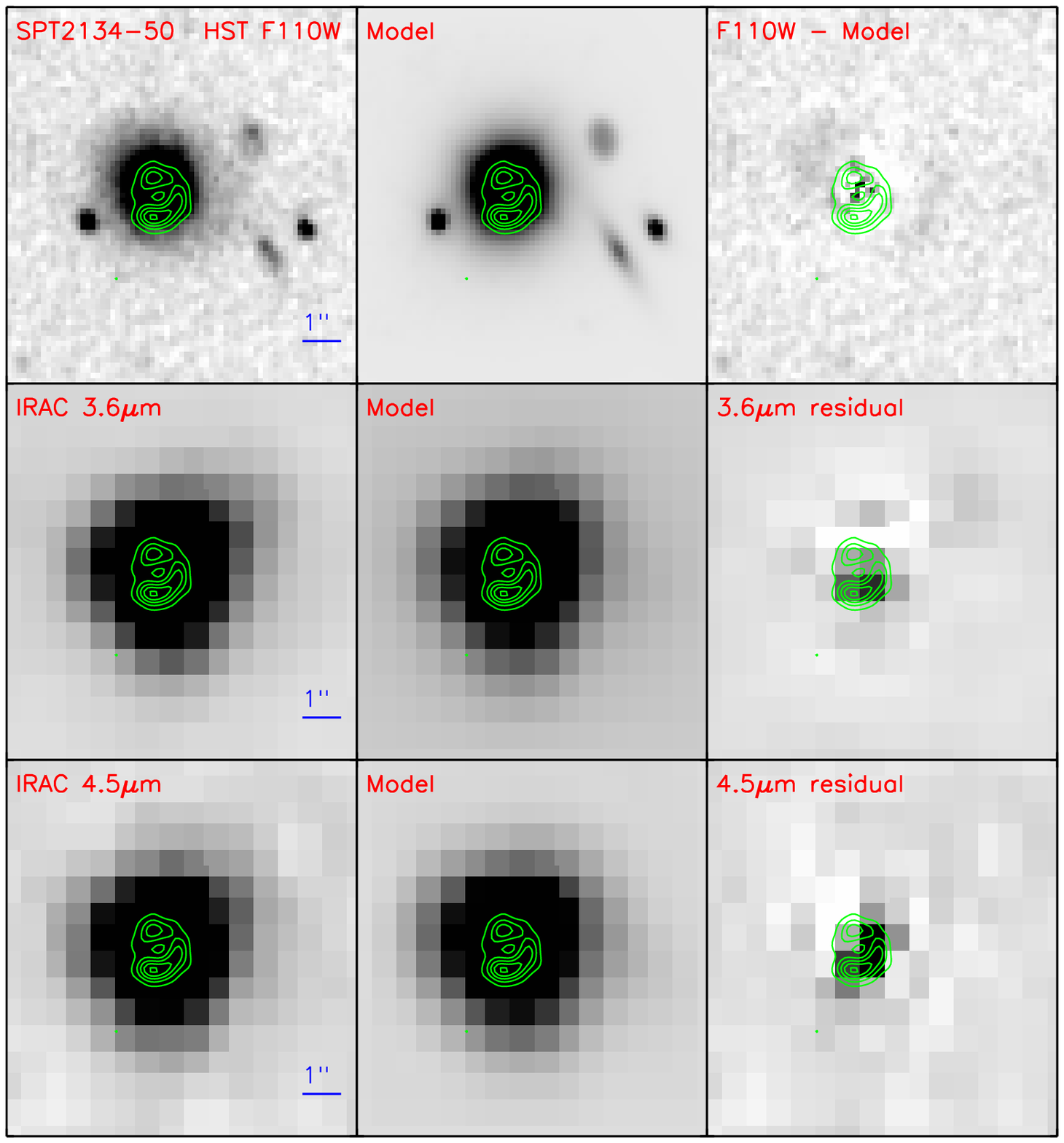}} 
\caption{9$\arcsec$ $\times$ 9$\arcsec$ {\it HST} and IRAC cutouts of SPT2134-50 showing the de-blending technique. Subfigures are the same as in Figure \ref{fig:0538} }
\label{fig:2134}
\end{figure*}

We simultaneously fit the lensing galaxy and 4 neighbors in the field (Fig.\ref{fig:2134} top panels). A point source was added to the central lens  in addition to the S\'ersic profile ($n$ = 1.9).  We see no trace of the emission suggested by the sub-millimeter ALMA contours in the residual image (top right).  Note that the lensing structure indicated by the ALMA contours is marginally resolved with a separation of 1.3$\arcsec$ in diameter, smaller than the separations in SPT0538-50 of 1.8$\arcsec$ at a similar redshift and SPT0103-45 of 2$\arcsec$ at even higher redshift. 

\subsubsection{SPT2146-55}

SPT2146-55 can be fit with a S\'ersic ($n$ = 3.1) plus an exponential disk in the F140W image. There is excess emission left over in the residual image, matching the ALMA contours. We find a detection of the background DSFG in this system at F140W with a flux density of $(1.5 \pm 0.4) \times 10^{-3}$ mJy.  A simple S\'ersic model also reveals the background source, albeit with worse central residuals. 

\begin{figure*}
\centering
{\includegraphics[width=17cm, height=15.5cm]{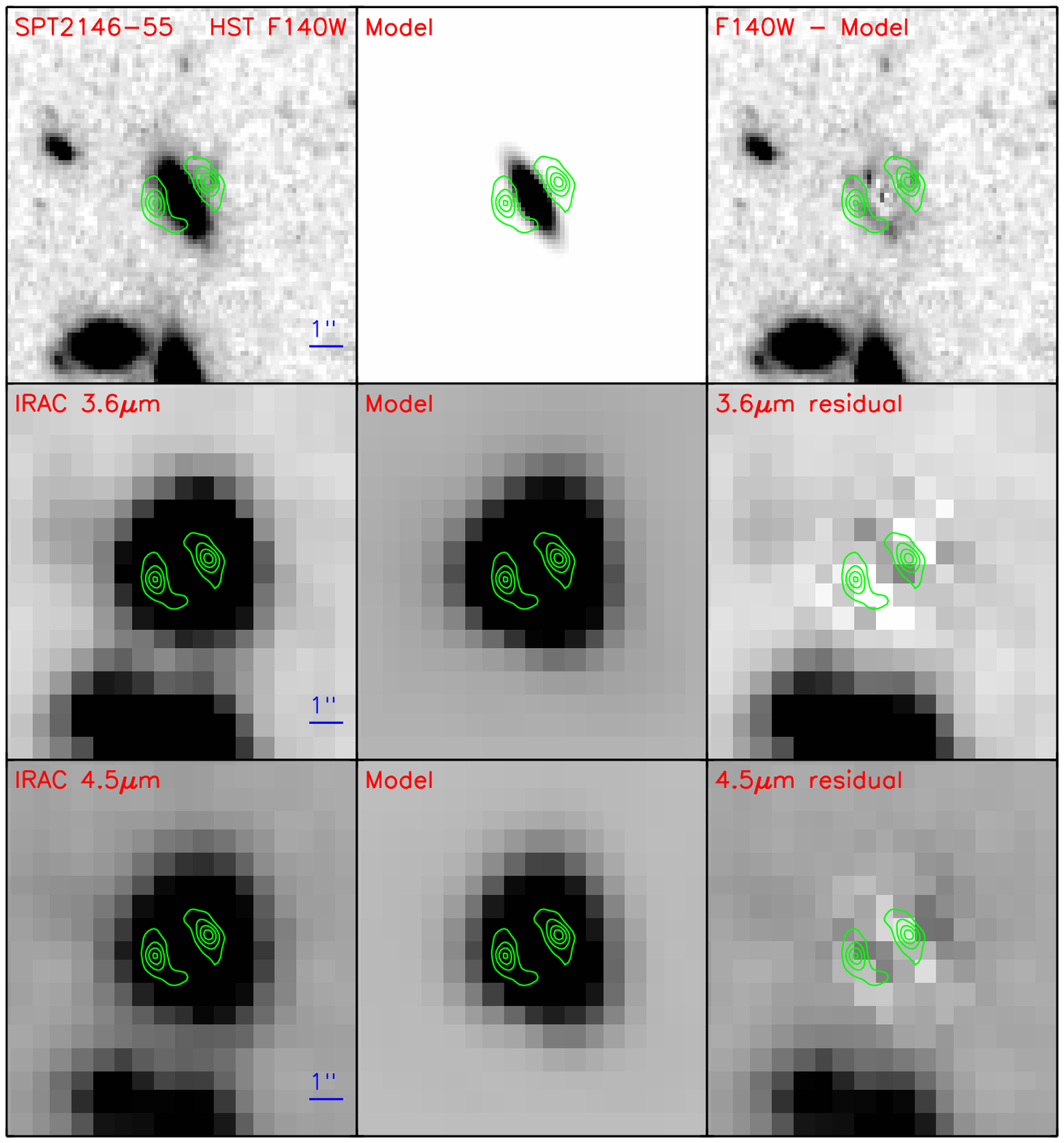}}  
\caption{10$\arcsec$ $\times$ 10$\arcsec$ {\it HST} and IRAC cutouts of SPT2146-55 showing the de-blending technique. Subfigures are the same as in Figure \ref{fig:0538}.  }
\label{fig:2146}
\end{figure*}

\subsubsection{SPT2147-50}

The lensing galaxy of SPT2147-50 is fit with a three-component SEP model of which the S\'ersic index is $n$ = 2.7. There is no residual associated with the ALMA contours.   \newline

\begin{figure*}
\centering
{\includegraphics[width=17cm, height=15.5cm]{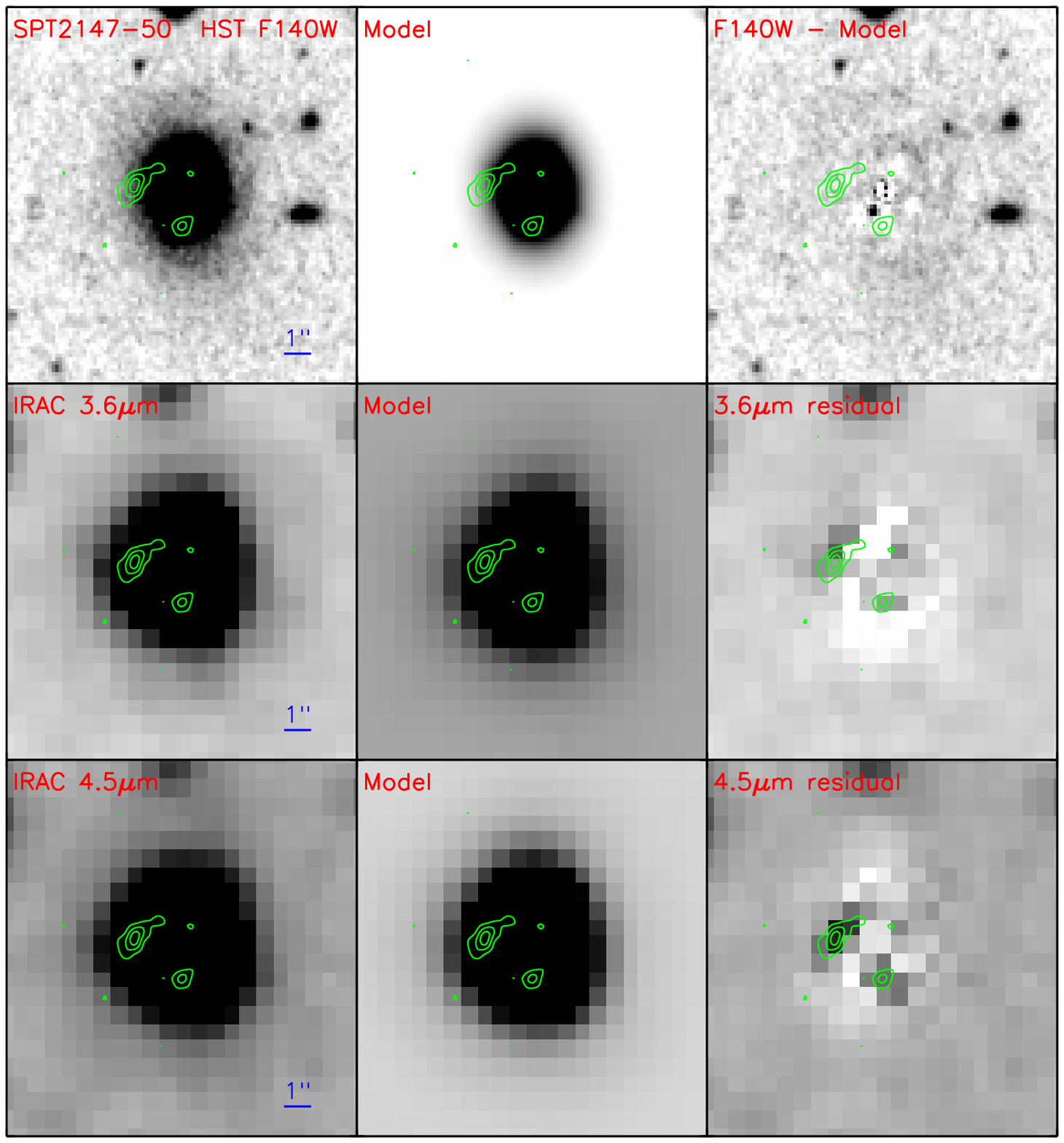}}  
\caption{13$\arcsec$ $\times$ 13$\arcsec${\it HST} and IRAC cutouts of SPT2147-50 showing the de-blending technique. Subfigures are the same as in Figure \ref{fig:0538}.  }
\label{fig:2147}
\end{figure*}

\subsection{De-blending the source/lens in IRAC}
\label{sec:deblendingIRAC}

\begin{figure*}
\centering
{\includegraphics[width=11cm, height=7cm]{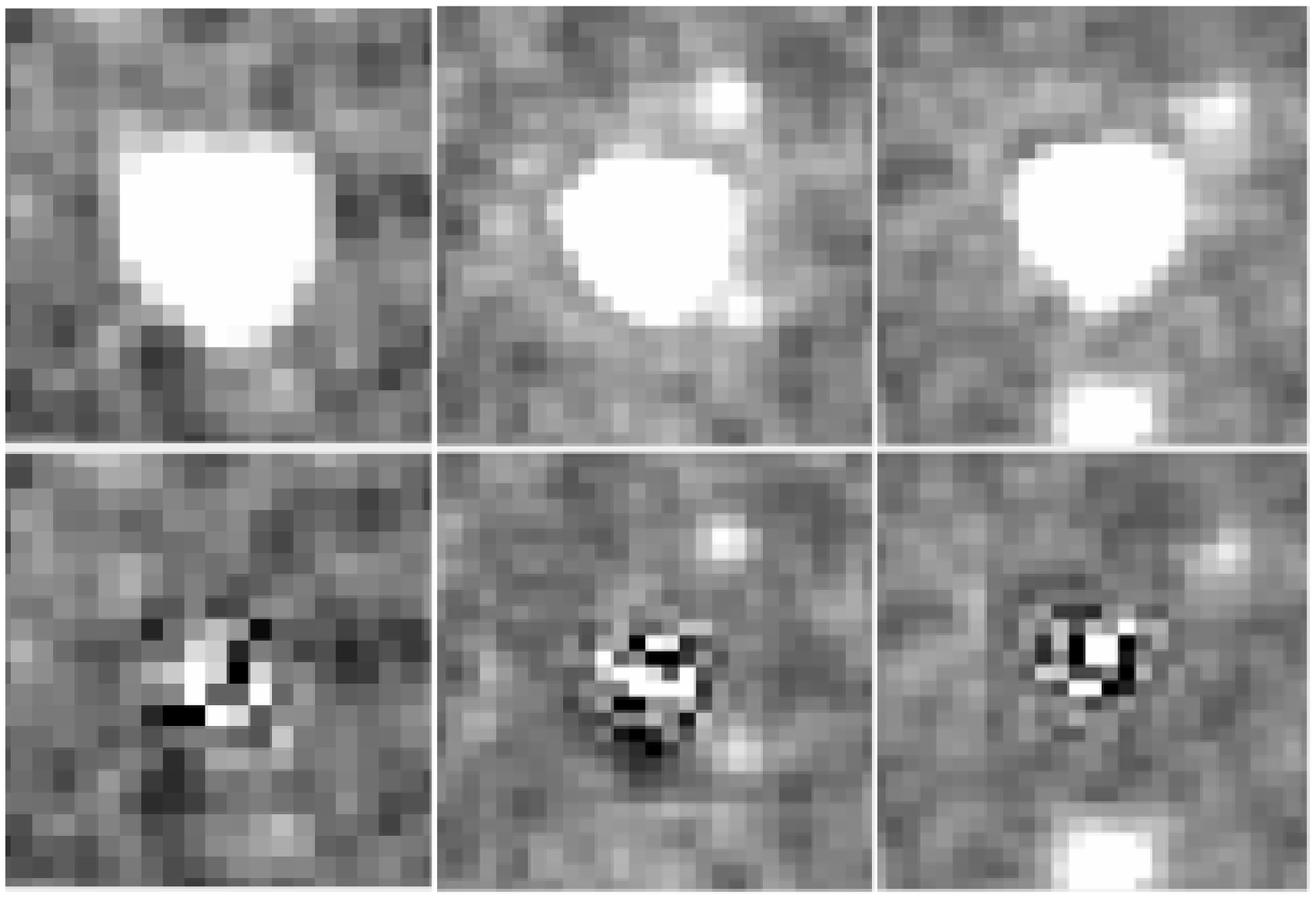}}
\caption{Examples of  stars in the field of  SPT0538-50 at 4.5 $\mu$m fit with the same IRAC PSF used in de-blending. {\it Top:} Original images. {\it Bottom:} Residual images after the original images subtracted by best-fit PSF.}
\label{fig:psf}
\end{figure*}

With the modeling of the lenses in the high-resolution {\it HST} images, we are able to disentangle the DSFG emission from the foreground emission in IRAC bands. The first attempt to disentangle the contributions from the foreground lens and the background source in low-resolution {\it Spitzer}/IRAC images of strongly lensed DSFGs was made by Hopwood et al. (2011). Similar de-blending techniques with additional refinements were implemented in subsequent works (e.g., \citealt{bothwell13b,Negrello14}).

We utilize the code Python Galaxy Fitter (PyGFit; \citealt{mancone13})  which is designed to measure PSF(point-spread function)-matched photometry from images with disparate pixel scales and PSF sizes. It takes models generated from a higher resolution image, {\it HST} in this case, and fits blended sources in crowded, low resolution images (i.e., {\it Spitzer}/IRAC). It should be noted that we make the assumption that the shape of the light profile of foreground galaxies remains the same from WFC3 1.1$\mu$m to IRAC 3.6$\mu$m, and 4.5$\mu$m, i.e. morphological {\it k}-corrections are negligible.  Before fitting the IRAC images by performing a $\chi^2$ minimization, the HST models are convolved with the IRAC PSF, which is empirically extracted by stacking at least five bright and isolated stars in the field. The quality of the IRAC PSF is critical to de-blending to avoid artifacts introduced by the PSF. We test the quality of the IRAC PSF by randomly selecting stars in the same field and fitting them with the same PSF used in de-blending.  We examine the star-subtracted residuals to assure the PSF is acceptable. A few examples of such fit and subtraction are shown in Fig. \ref{fig:psf}. The residual flux from these stars is less than 1\% of the original flux in all cases.

We optimize the fitting by masking out the region where the DSFG emission is expected based upon the ALMA data, thereby fitting only the foreground lens. During the $\chi^2$ minimization only the positions and fluxes of the objects are left as free parameters. All other S\'ersic parameters (effective radius, S\'ersic index, aspect ratio, and position angle) are held fixed. The positions are restricted to small shifts (typically less than a pixel).  We examine the residuals to see if there is excess emission associated with the ALMA contours. 

\subsubsection{SPT0538-50}
\label{sec:0538irac}

The middle row and bottom row in Fig. \ref{fig:0538} show the de-convolution process just described at 3.6$\mu$m and 4.5$\mu$m adopting the SEP model. The best-fit IRAC model ({\it Middle}), constructed by convolving the {\it HST}/WFC3 model with the IRAC PSF, is removed from the original 3.6$\mu$m/4.5$\mu$m image ({\it Left})  to produce the residual image on the right.  The Einstein ring is clearly recovered in both IRAC bands, closely matching the ALMA contours in green.  We perform aperture photometry on the IRAC residual images, using an annular aperture large enough to enclose the 3$\sigma$ ALMA contour without including too much sky flux.  
The resultant sky-subtracted flux densities (Table \ref{tab:photometry}) at 3.6$\mu$m  and 4.5$\mu$m are $S_{3.6}$ = $26.2\pm6.7 \mu$Jy and $S_{4.5}$ = $53.7\pm7.0 \mu$Jy. We note that these are consistent with the values of $S_{3.6}$ = $22 \pm 5 \mu$Jy and $S_{4.5}$ = $47 \pm 8 \mu$Jy derived by \cite{bothwell13b}.  The uncertainties include the photometric uncertainties as well as the uncertainty associated with the de-blending process where the residual fluxes can vary due to a different set of PyGFit input parameters. We find that the maximum allowed positional shift during the $\chi^2$ fitting dominates the uncertainties in the resultant residual structures, and therefore the residual fluxes particularly for simultaneous fitting of the source consisting of multi-components. The above best-fit model is determined by $\chi^2$ minimization. We apply the $\Delta\chi^2$ technique to determine the $1\sigma$ error by varying the input positions in PyGFit. We change the allowed maximum positional shift during fitting from 0 to 0.6$\arcsec$ (1 pixel) until $\chi^2$ increases by 2.3. 

\subsubsection{SPT0103-45}

We apply the analogous de-blending technique to SPT0103-45 and it is detected in both bands, with excess emission coincident with the ALMA contours. We put an annular aperture on the residual images, extracting the flux densities at two bands of $S_{3.6}$= $5.1 \pm 1.3 \mu$Jy and $S_{4.5}$= $12.8 \pm 1.4 \mu$Jy. The errors are determined in the similar way. 

\subsubsection{SPT0346-52}

The middle row and bottom row in Fig. \ref{fig:0346} show the results of the de-convolution process for SPT0346-52 in IRAC bands.  The 3.6$\mu$m and 4.5$\mu$m images are subtracted clean. SPT0346-52 is a non-detection in IRAC with 3 hour exposure at 4.5 $\mu$m even with the help of strong gravitational lensing. Formally we measure the flux to be 0.5 $\pm$ 0.8 $\mu$Jy at 3.6$\mu$m and 0.3 $\pm$ 1.2 $\mu$Jy at 4.5$\mu$m. We use the non-detections in both IRAC bands to put upper limits on the fluxes from SPT0346-52 based on the sky flux distribution (random aperture) in the residual maps.  We place a 3.6$\mu$m band 3$\sigma$ upper limit of  2.4  $\mu$Jy  and a 4.5$\mu$m band 3$\sigma$ upper limit of  3.6 $\mu$Jy .

\subsubsection{SPT2134-50}

Examining the IRAC residual maps (Fig. \ref{fig:2134}) of SPT2134-50, we find the excess emission coincident with the ALMA contours. However, due to the small separation of the lensing arcs, imperfections in the {\it HST} model could also contribute to the residual in IRAC given the relatively large IRAC PSF.  In this case, we failed to extract robust fluxes from the residual because the separation was not large enough ($\sim 1.7\arcsec$ in diameter) to avoid confusion. We instead extracted the fluxes as 3$\sigma$ upper limits of  $<20.0$ $\mu$Jy at 3.6$\mu$m and  $<32.9$ $\mu$Jy at 4.5$\mu$m, providing a loose constraint on the stellar mass. 

\subsubsection{SPT2146-55}

With the detection in HST F140W, we expected to detect SPT2146-55 in the IRAC bands. However, the 1 hour IRAC data do not provide clear detections, given that the residuals do not match the ALMA contours and are not clearly related to the background source. The measured fluxes are 0.5 $\pm$ 1.4 $\mu$Jy at 3.6$\mu$m and 0.4 $\pm$ 2.3 $\mu$Jy at 4.5$\mu$m. We instead adopt 3$\sigma$ upper limits estimated from the background sky using a ring aperture: $<$ 5.2 $\mu$Jy at 3.6 $\mu$m and $<$ 6.9 $\mu$Jy at 4.5 $\mu$m. 

\subsubsection{SPT2147-50}

After the de-blending process for SPT2147-50, there are three blobs coincident with the ALMA contours. We use separate apertures corresponding to the separate ALMA contours to extract fluxes instead of an annular aperture to avoid the over-subtracted regions. We obtain flux densities of 2.0 $\pm$ 0.6 $\mu$Jy at 3.6 $\mu$m and 3.5 $\pm$ 0.4 $\mu$Jy at 4.5 $\mu$m.

\begin{table*}
\centering
\caption{Multi-band observed flux densities in mJy}
\begin{tabular}{@{}lcccccc@{}}
\hline
\hline
Observed Wavelength $ \lambda $&  SPT0103-45  &  SPT0346-52& SPT0538-50 &SPT2134-50&SPT2146-55  &SPT2147-50\\
 \hline
 {\it HST}/WFC3  1.1$\mu$m	& $<5.8 \times 10^{-4} $ &   	 $< 3.8 \times 10^{-4}$ & $<$ 0.0022 &                        $<$ 0.0012& &\\
 {\it HST}/WFC3  1.4$\mu$m    	&  & & & & $(1.5 \pm 0.4) \times 10^{-3} $ &     $< 7.3  \times 10^{-4} $ \\
{\it HST}/WFC3  1.6$\mu$m	& $< 0.0026 $ &                            $<9.1 \times 10^{-4} $ 	& $<$ 0.0033 &                   $<$ 0.0016 & &\\
{\it Spitzer}/IRAC 3.6$\mu$m	& $0.0051 \pm 0.0013 $ &                            $<$ 0.0024	& $0.0262 \pm 0.0053$ &            $<$ 0.0200   &$<$ 0.0052 &                   $0.0020 \pm 0.0006  $\\
{\it Spitzer}/IRAC 4.5$\mu$m	& $0.0128 \pm 0.0014$ &                             $<$ 0.0036  & $0.0537 \pm 0.0070$ &        $<$ 0.0329	& $<$ 0.0069 &                   $0.0035 \pm 0.0004  $ \\
{\it Herschel}/PACS100$\mu$m	&     $<13$        &                             $<6$ & $31 \pm 2$  &                  $49 \pm 3$    & $< 8 $ &   $9 \pm 2$ \\
{\it Herschel}/PACS160$\mu$m	&    $<47$         &                             $33 \pm 9$& $141 \pm 15$&                $196 \pm 22$& $< 29 $&  $< 28$\\
{\it Herschel}/SPIRE 250$\mu$m	&$133 \pm 14$ & 			$122 \pm 11$ &$326 \pm 23$ &  	    $350 \pm 25$&$65 \pm 13$ & $72 \pm 9$ \\
{\it Herschel}/SPIRE 350$\mu$m	& $213 \pm 16$& 			$181 \pm 14$	& $396 \pm 38$&		  $332 \pm 23$& $69 \pm 13$& $115 \pm 10$\\
{\it Herschel}/SPIRE 500$\mu$m	& $232 \pm 17$&			 $204 \pm 15$& $325 \pm 24$& 		    $269 \pm 19$	 & $83 \pm 10$&  $121 \pm 11$\\
APEX/LABOCA 870$\mu$m 	& $125 \pm 6$&					 $131 \pm 8$	& $125 \pm 7$&	   $101 \pm 7$ 	& $55 \pm 4$&  $61 \pm 5$\\
SPT 1.4mm          & $39.1 \pm 7.0$ &							$46.0 \pm 6.8$   & $28 \pm 4.6$ &		   $22.8 \pm 4.6$ & $17.8 \pm 3.9$ &  $20.3 \pm 4.6$ \\
SPT  2.0mm         & $8.8 \pm 1.4$ &							 $11.6 \pm 1.3$      & $8.5 \pm 1.4$ &       $6.0 \pm 1.3$ & $5.2 \pm 1.3 $ & $5.8 \pm 1.3$ \\
\hline
\end{tabular}
\tablecomments{All the {\it HST} and IRAC photometry is derived in this work. The quoted uncertainties include photometric uncertainties plus uncertainties due to the de-blending process. For the non-detections, the flux density upper limits are given at 3$\sigma$. The FIR photometry is from Strandet et al. in prep except SPT0538-50 for which we use the photometry by \cite{bothwell13b}. }
\label{tab:photometry}
\end{table*}

\section{Lens Models}
\label{sec:lensmodel}

The lens modeling technique, based on our Cycle 0 ALMA Band 7 imaging,  is described by \cite{hezaveh13}. In this paper,  six gravitationally lensed SPT sources, including SPT0538-50 and SPT0346-52, are discussed in detail.  To summarize, the modeling is performed in the visibility plane to properly compare with data as the ALMA interferometer samples the Fourier transform  of the sky brightness distribution across a 2-D spatial frequencies instead of directly imaging the emission.  The source is modeled as a symmetric Gaussian or a S\'ersic light profile while the lens is assumed to be a Singular Isothermal Ellipsoid (SIE).  Minimizing $\chi^2$ between data and model visibilities determines the best-fit. The lensing magnification factor, $\mu$, is defined as the ratio of the total lensed to unlensed flux. The lens models allow us to derive intrinsic properties of the lensed galaxies by simply dividing the observed properties by $\mu$. We present the resultant magnifications for the sources discussed in this work in Table \ref{tab:params}. We adopt the latest magnification factors for all the six sources derived in the same fashion in Spilker et al.  (in preparation). When deriving stellar masses and SFRs, we correct for the lensing magnifications and propagate the corresponding errors. We consider the issue of differential magnification in Section \ref{sec:ssfr}.

\section{SED fitting}
\label{sec:SEDfitting}

To derive properties of these dusty star-forming galaxies such as stellar mass, infrared luminosity ($L_{IR}$  from 8\---1000 $\mu$m), and SFR, we utilize the code CIGALE \citep{burgarella05, noll09}\footnote{http://cigale.lam.fr}  to fit our photometric data from near-IR to far-IR.  For the fitting, the code generates FUV to FIR SEDs consisting of dust-attenuated complex stellar population models, IR dust emission models, and spectral line templates.  It allows self-consistent treatment of the stellar component, sampled by our NIR data and the re-emitted dust component, sampled by the FIR data. We adopt the \citeauthor{maraston05} (2005; M05) stellar population synthesis model, a \cite{kroupa01} initial mass function (IMF),  and a solar metallicity as our baseline.

\subsection{Star formation histories}

One- or two-component star formation history (SFH) models with exponentially increasing or decreasing SFR or continuous constant SFR  are possible to define the star formation history in CIGALE.  
Commonly assumed star formation histories in the literature are: (1) a single stellar population with an exponentially decreasing SFR ($\tau$ model; e.g., \citealt{Ilbert09,Ilbert10,Ilbert13,Muzzin13}), (2) an exponentially increasing SFR (e.g., \citealt{Maraston10,Papovich11}), (3) delayed $\tau$ model (e.g. \citealt{dacunha15}), (4) an instantaneous burst of star formation or constant star formation (e.g. \citealt{kennicutt98,murphy11}), and (5) a model with two stellar populations: a young stellar population with a constant SFR on top of an old stellar population with an exponentially declining SFR (e.g., \citealt{Papovich01,Lee09}).  

\subsubsection{Single-component SFH model}

The most commonly used SFH model is the exponential or $\tau$-model with the simple form of SFR $\propto$ $e^{-t/\tau}$. This model naturally arises in scenarios where the SFR is directly proportional to the gas density in a closed-box model \citep{Schmidt59}. There are two parameters, the e-folding time $\tau$ and age which we consider as free parameters. If $\tau$ $>>$ age, the $\tau$-model transforms to a single burst star formation history with a constant rate.  This simple model is widely adopted in the literature, facilitating comparison with previous work (e.g., \citealt{karim11,Heinis14,Bethermin15a}).

\subsubsection{Two-component SFH model}
\begin{figure}
\centering
{\includegraphics[width=9cm, height=11cm]{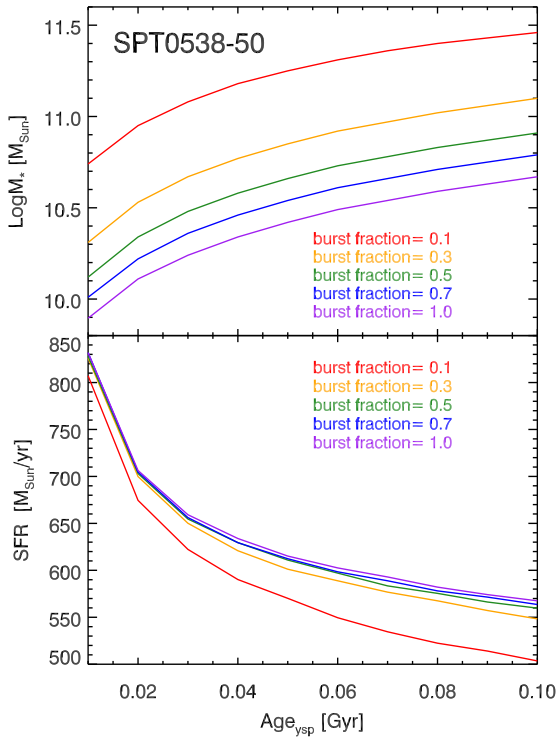}}      
\caption{The dependence of derived stellar mass on the age of the young stellar population and the burst fraction for SPT0538-50.}
\label{fig:0538ysp}
\end{figure}

We also investigate the two-component SFH model because real systems are expected to experience multiple episodes of star formation \citep{Papovich01,Lee09}. Moreover, the double SFHs are likely to better reproduce the true stellar mass \citep{michalowski14}. The two stellar population components are connected by the burst strength/fraction $f_{ysp}$ (i.e., the mass fraction of young stellar population in the model). We first treat the ages of both the young and old stellar populations and the burst fraction as well as Av as free parameters. The derived instantaneous SFRs and stellar masses are poorly constrained. These two parameters, the age of the young stellar population and the burst fraction, are the dominant ones that drive the resultant stellar mass. The dependence of derived stellar mass and SFR on varying ages and burst fractions is investigated. We show the case of SPT0538-50 as a demonstration in Figure \ref{fig:0538ysp}. The other SPT DSFGs exhibit similar trends. We allow the age of the young stellar population to vary from 10 to 100 Myr in steps of 10 Myr (commonly adopted starburst lengths $\sim$10-100 Myr, e.g., \citealt{michalowski10,murphy11,kennicutt98}). We permit the burst fraction to vary from 0 to 1, noting that burst fractions of $<$ 0.1 are ruled out at high confidence. We generate a grid of models of different input $f_{ysp}$ and $t_{ysp}$. The resulting SEDs are equally well fit in terms of matching the IRAC and FIR data points and being consistent with the {\it HST} upper limits.  As a result, the stellar mass increases by $\sim$0.8 dex with an increasing age from 10 to 100 Myr. It is also subject to the burst fraction at a fixed age: $M_*$ drops by $\sim$ 0.8 dex from $f_{ysp}$ = 0.1 to 1. The instantaneous SFR is much less dependent on $f_{ysp}$ than it is on $t_{ysp}$. This suggests that we do not have enough leverage to constrain these key parameters and the stellar mass heavily depends on the assumptions made for the two parameters.

We demonstrate in the SFR -- M$_*$ plane (Fig. \ref{fig:SFRmstar})  that SFRs modeled by the single-component SFH are better constrained than those modeled by the double-SFH. We adopt the simplest single-component SFH model hereafter since we do not have enough data to constrain a more complex SFH. We present the SED fitting results in the next section with the $\tau$-model adopted.

\subsection{SED fitting results: Stellar mass and SFR}
\label{sec:fittingresults}

We run CIGALE, with the input parameters listed in Table \ref{tab:SEDparameter}, to find the best-fit SED and inferred stellar mass,  IR luminosity, and instantaneous star formation rate.  Expectation values and standard deviations for the output galaxy properties are derived from the probability distribution function of the parameter-bin-specific best-fit models. Fig. \ref{fig:SED} shows the individual best-fit SEDs for the six SPT DSFGs in the observed frame.  We also obtain an intrinsic (i.e., magnification corrected) average SPT SED (in black in Fig. \ref{fig:composite}) by taking the median value of the six best-fit SEDs at each rest-frame wavelength. We report stellar masses, infrared luminosities, SFRs and associated uncertainties in Table \ref{tab:stellarmass}. The FIR part of the SED is well constrained, thus the IR luminosity has little dependence on varying input parameters. The optical/UV part is less constrained, therefore the derived stellar mass depends strongly on the input parameters.  The dust extinction is a free parameter in the fitting and we find a significant attenuation by dust ($A_V \sim 4 - 5$) is required to match the IRAC photometry and the FIR data points, which is consistent with other SMGs (e.g., \citealt{negrello10,hopwood11,hainline11,michalowski10}). The derived ages are all much smaller than the corresponding e-folding times suggesting that the star formation history essentially can be described by a single extended burst star formation. The best-fit burst ages are 20 -- 80 Myr. However, we note that there can exist older stellar populations that are hidden by the bright young population (\citealt{Maraston10, Buat14}).

Our stellar mass estimates for the six SPT DSFGs span more than an order of magnitude, with a median value around  $5 \times 10^{10}M_{\Sun}$, for those detected. The median stellar mass is  lower than the median values of a well-studied SMG sample \citep{chapman05} found by \citeauthor{michalowski10} (2010; $3.5 \times10^{11}M_{\Sun}$) and the recent ALESS sample by \citeauthor{dacunha15} (2015; $8.9 \times10^{10}M_{\Sun}$). These two studies perform multi-band SED fitting from rest-frame UV to FIR as we do. However, these two samples are biased towards higher stellar masses while ours are biased towards lensing configuration. 

All of the SPT DSFGs present prodigious star formation at rates $>$ 500 $M_{\Sun}{\rm yr}^{-1}$.  SPT0346-52 stands out as having a SFR of $\sim$4800 $M_{\Sun}{\rm yr}^{-1}$ only $\sim 1$ billion years after the Big Bang, among the highest SFRs at any epoch.  Given that the $L_{IR}$ is the best measured quantity and is widely used as an SFR indicator, we also calculate the SFRs using $L_{IR}$ conversion recipes of \cite{murphy11} and \cite{kennicutt98}  with the common IMF of Kroupa and continuous bursts of age 10-100 Myr.  When converting SFR from Salpeter IMF to Kroupa IMF, we divide the conversion coefficient by 1.53. Both estimates are presented in Table \ref{tab:stellarmass}. All of the SFRs derived by CIGALE SED fitting are consistent with those by the conversion recipes.

\subsubsection{Systematics}

A scatter of $\sim$ 0.3 dex around the correct value is an intrinsic uncertainty in determining stellar masses from broad-band photometry through SED fitting \citep{michalowski14}. For our SPT DSFGs, we have checked that stellar masses derived using the Bruzual $\&$ Charlot (2003; BC03) models are at most 0.1 dex higher than masses derived using the M05 models\footnote{The SED fitting in this work is mainly based on CIGALE FORTRAN where BC03 model is not included. BC03 model is recently available in the CIGALE python version at http://cigale.lam.fr/pcigale/index\_pcigale.html. We use this new version in testing the discrepancy caused by choice of synthesis models.}.  A two-component SFH on average yields stellar masses $\sim$ 0.15 dex higher than a single-component SFH. These systematic uncertainties are subdominant compared to the statistical uncertainties ($\sim$ 0.4 dex) for our data. Our uncertainties are dominated by the limited number of photometric data points at optical/near-IR wavelengths.

\begin{figure*}
\centering
{\includegraphics[width=10cm, height=7cm]{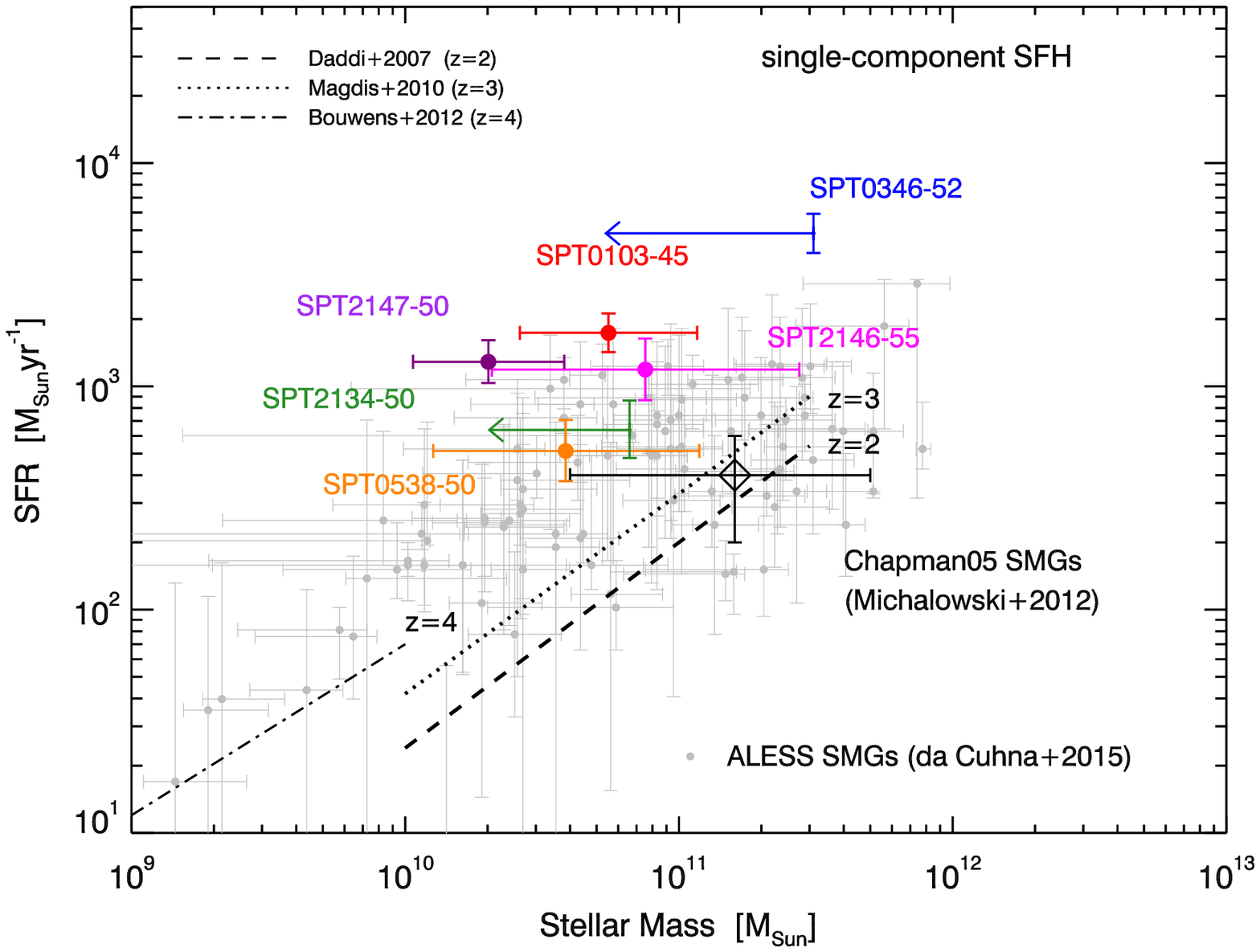}}  
{\includegraphics[width=10cm, height=7cm]{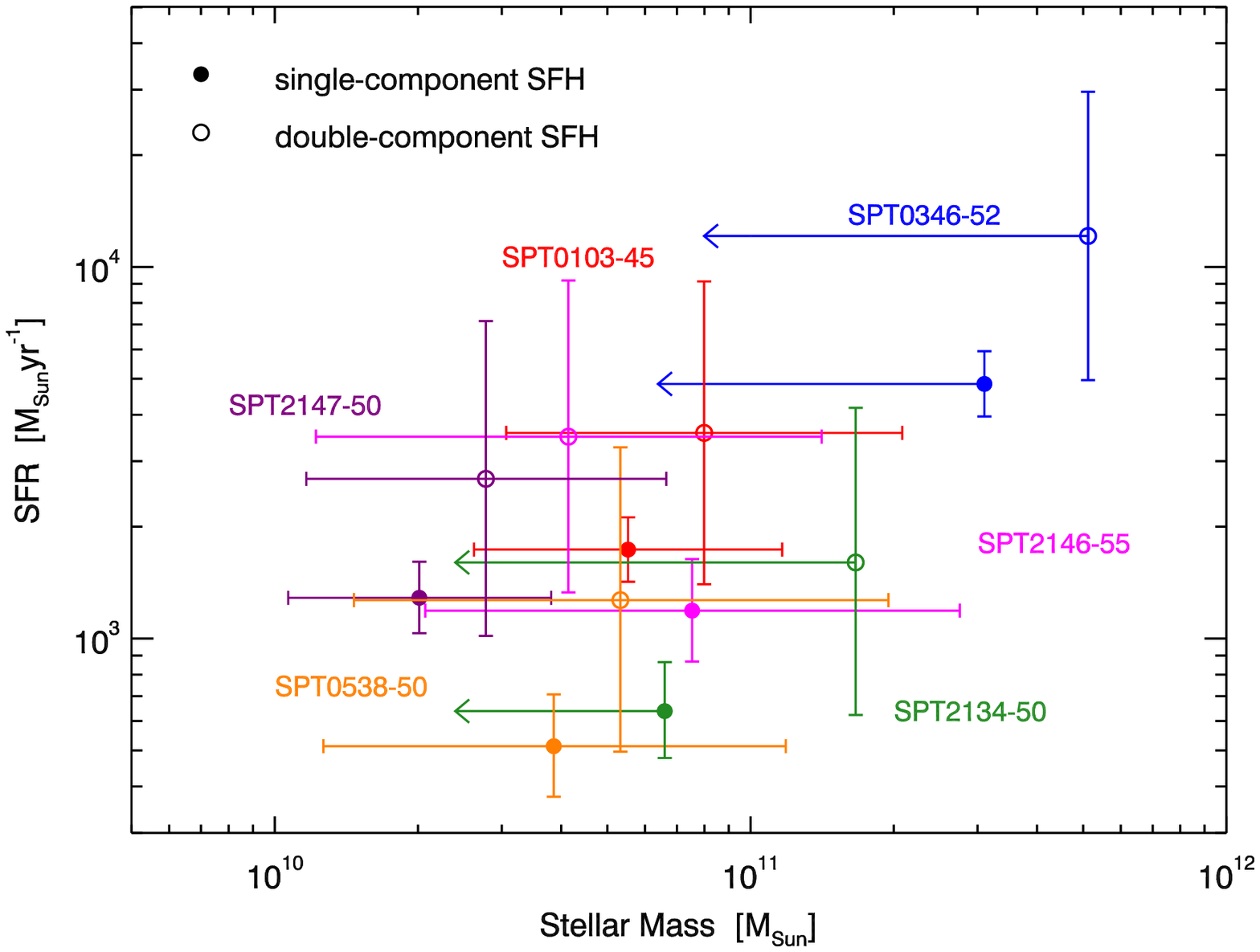}}  
\caption{{\it Top:} SFR--M$_*$ relation of SPT DSFGs derived from CIGALE SED fitting using the single component SFH. The error bars are derived from the probability distribution function by CIGALE.  The dashed line shows the main sequence of star-forming galaxies at $z$ = 2 (\citealt{daddi07}), the dotted line is the relation at $z$ = 3 from \cite{magdis10}, and the dash-dotted line shows the relation at $z$ =4 at lower stellar masses by \cite{bouwens12}. The median $SFR-M_*$ relation of  \cite{chapman05} SMGs at $z$ = 2-3 is shown by the open diamond. These SMGs are extensively explored in \cite{michalowski12} by assuming different star formation histories which are reflected in the error bars.  {\it Bottom:} The comparison between the stellar masses and SFRs derived using a single-component SFH and a double-component SFH.}
\label{fig:SFRmstar}
\end{figure*}

\begin{figure*}
\epsscale{0.5}
\plotone{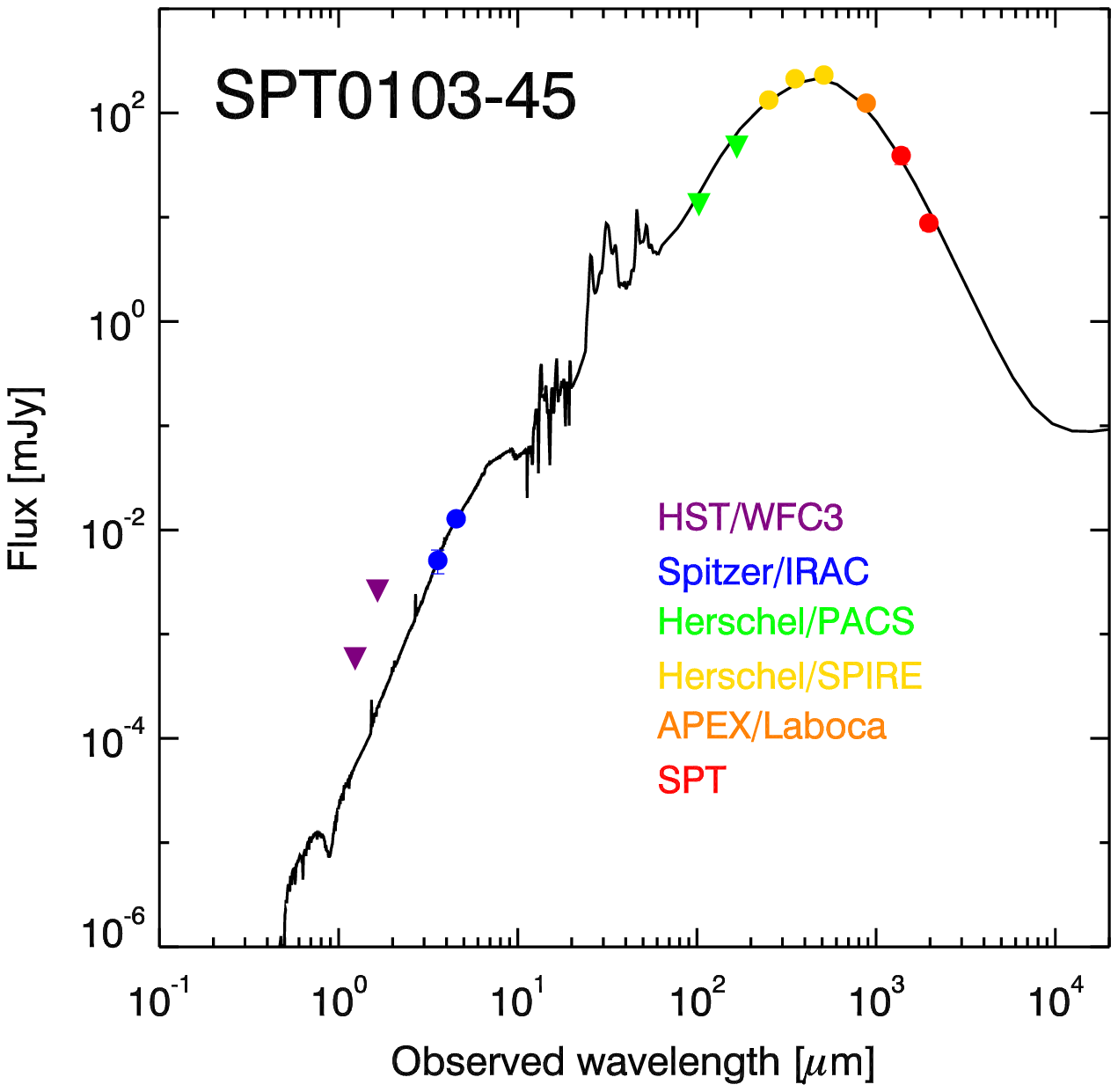}  
\plotone{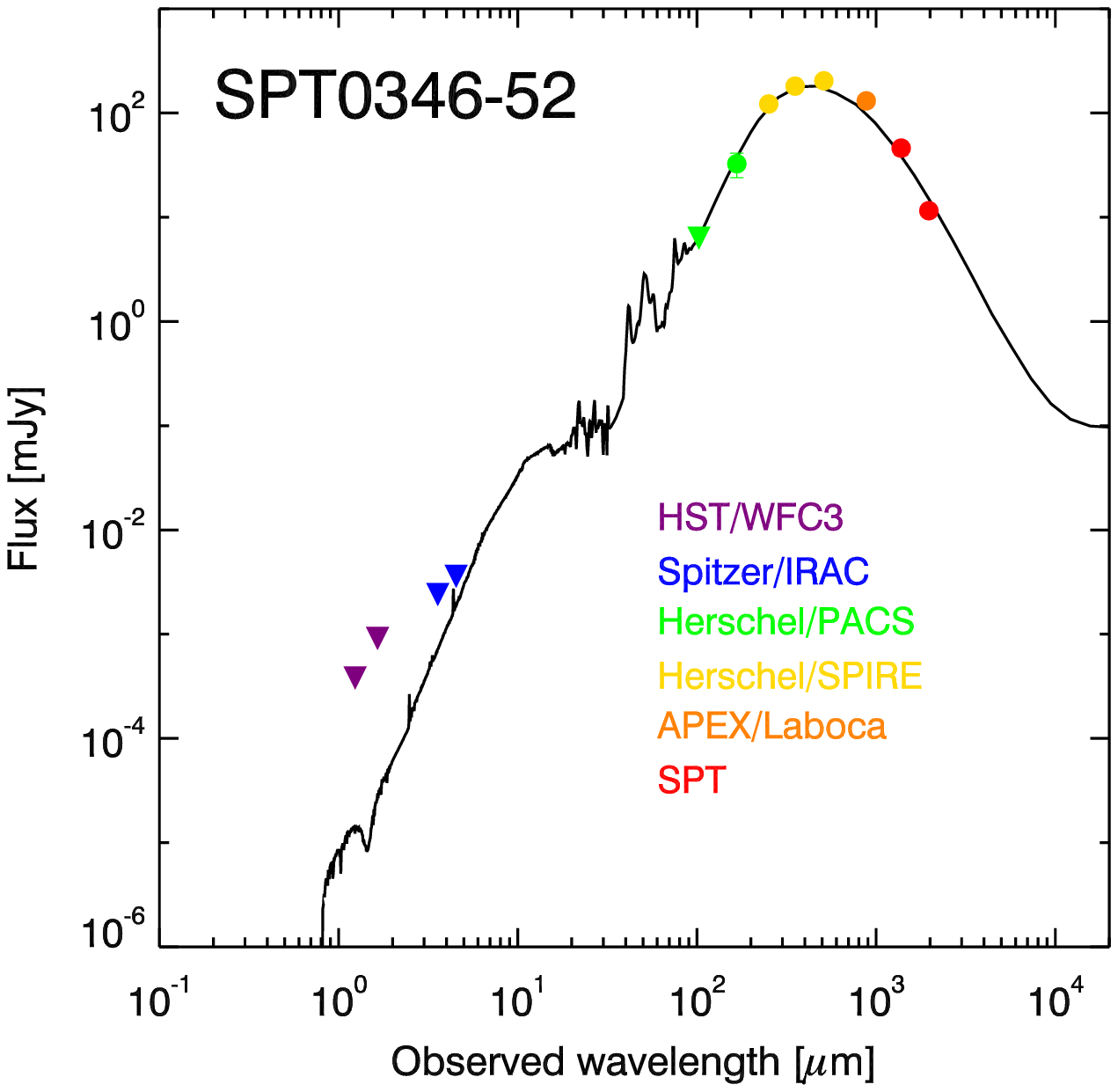}     
\plotone{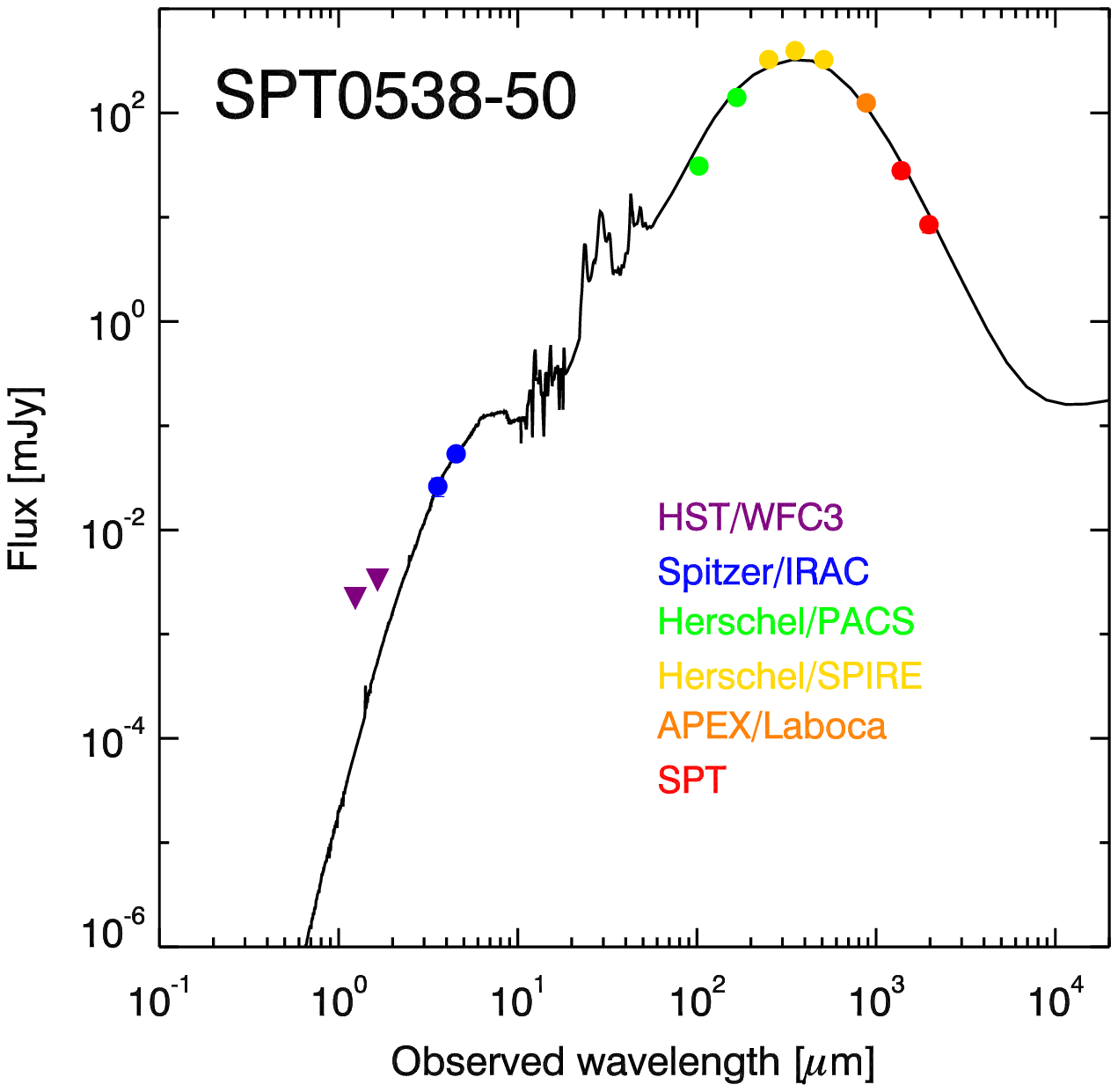}     
\plotone{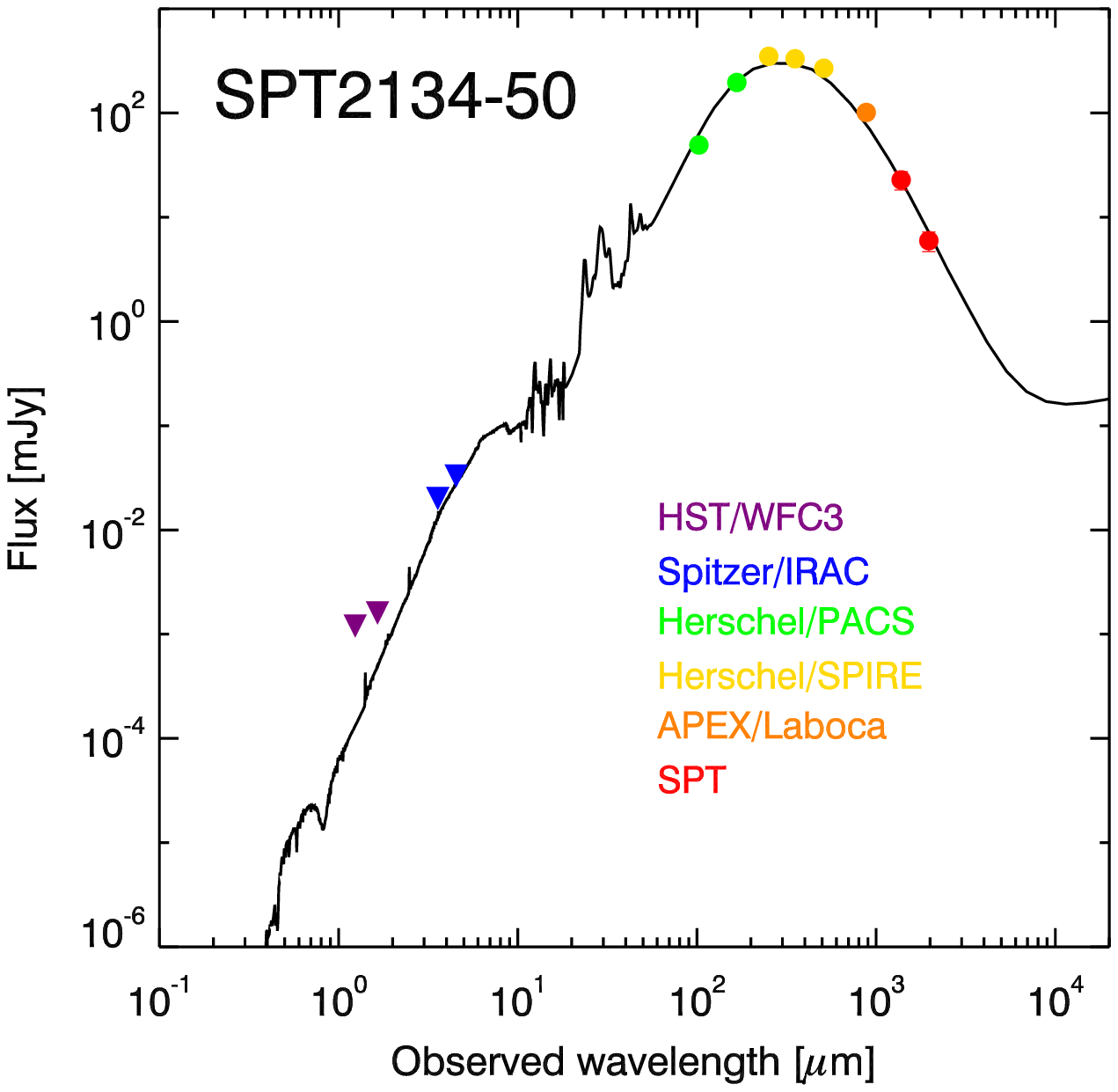}    
\plotone{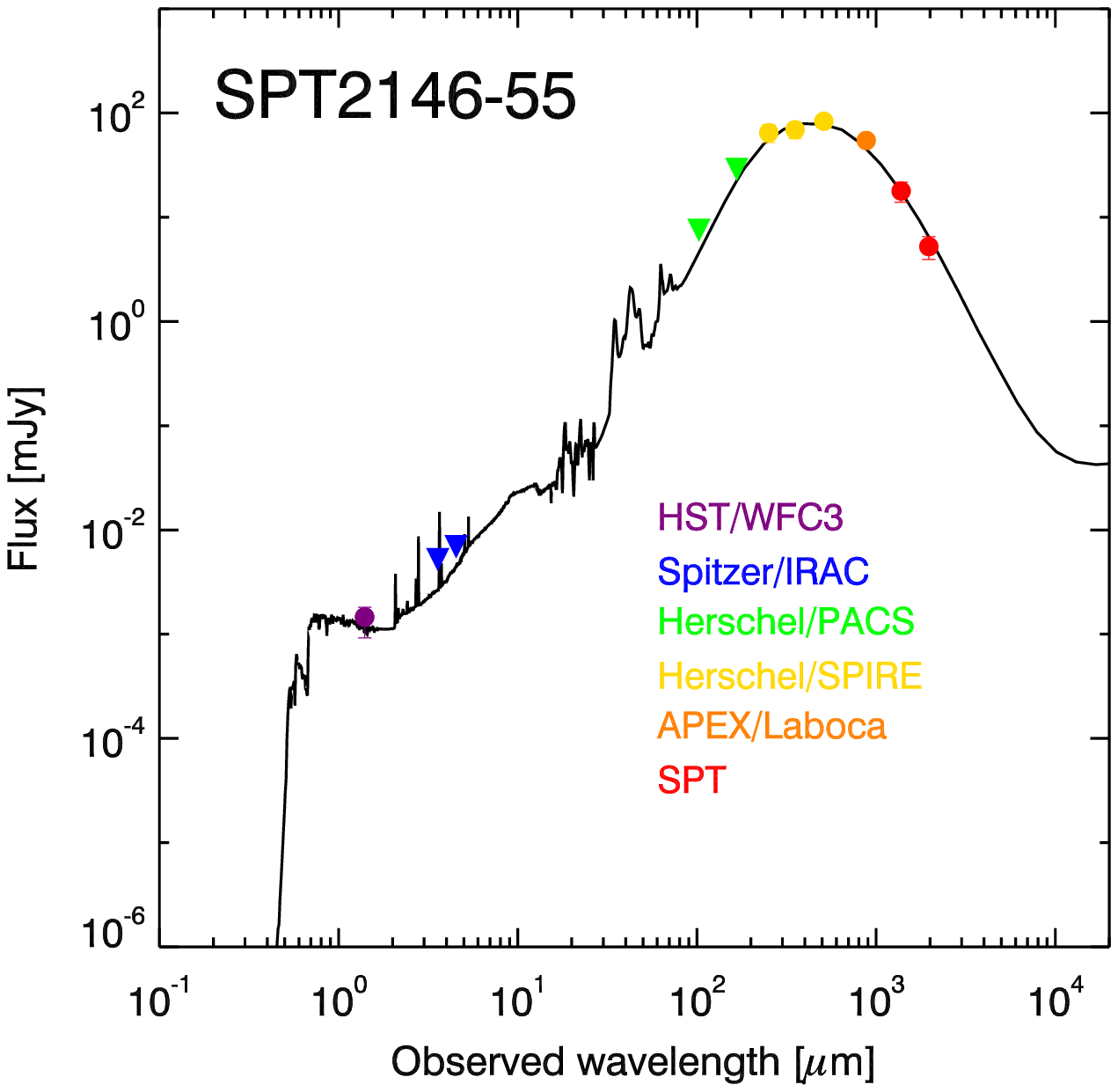}    
\plotone{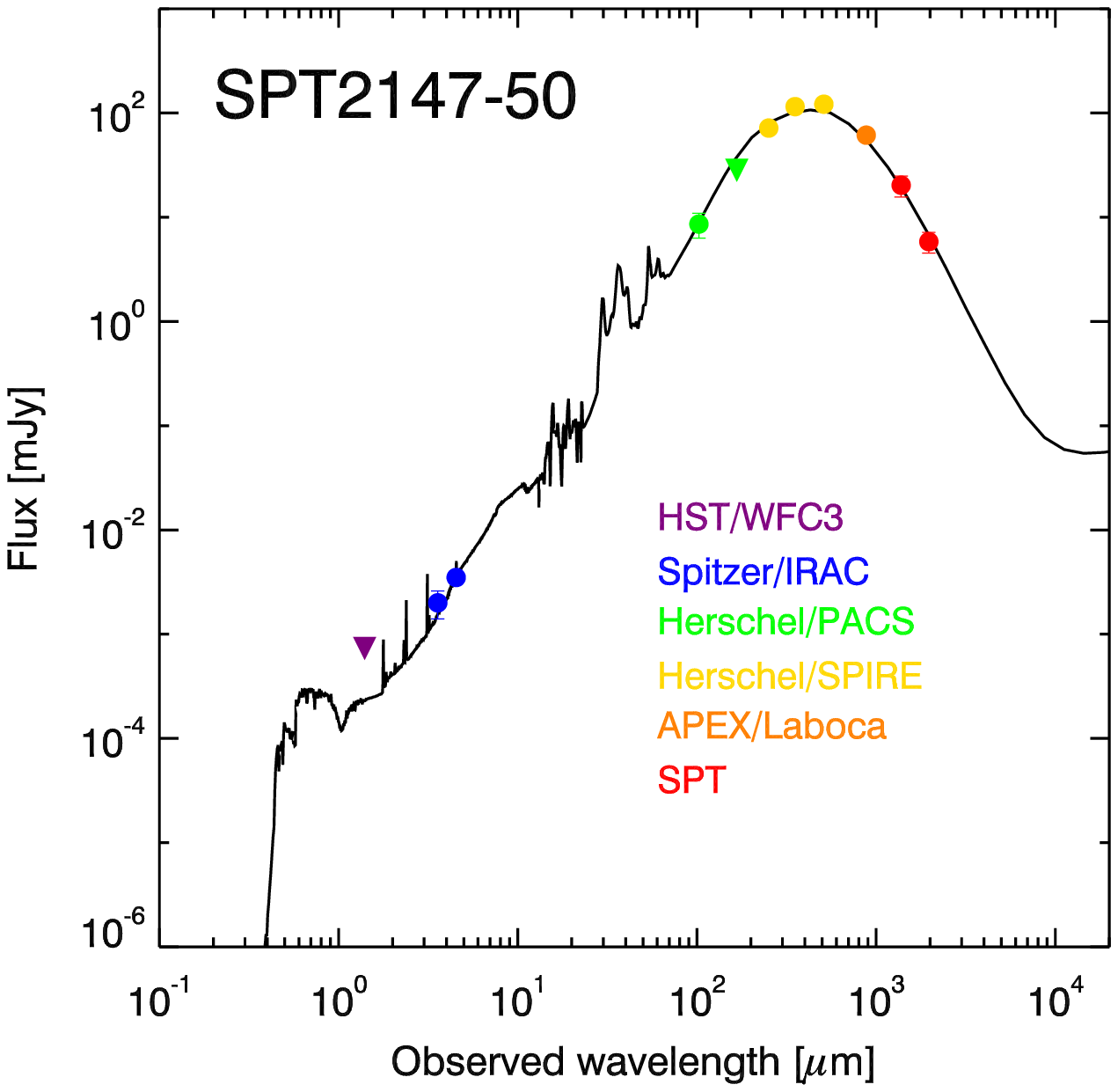}   
\caption{The SED fitting results for the six SPT DSFGs using the SED fitting code CIGALE, based on the assumptions of the M05 stellar population synthesis models and a single-component SFH. The data points from left to right are HST/WFC3 F110W+F160W/F140W, Spitzer/IRAC 3.5$\mu$m+4.6$\mu$m, Herschel/PACS 100$\mu$m+160$\mu$m, Herschel/SPIRE 250$\mu$m+350$\mu$m+500$\mu$m, APEX/LABOCA 87$\mu$m, and SPT 1.4mm+2.0mm. The circles denote detections while the triangles are 3$\sigma$ upper limits.}
\label{fig:SED}
\end{figure*}

\begin{figure}
\centering
{\includegraphics[width=9cm, height=7cm]{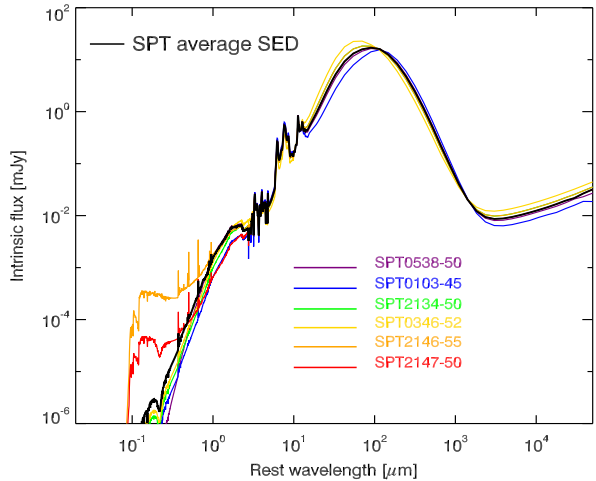}} 
\caption{An intrinsic average SED (black) obtained by taking the median value of the six best-fit SEDs (color-coded) at each rest-frame wavelength. They are normalized to the total infrared luminosity of the average SED. }
\label{fig:composite}
\end{figure}

\begin{table*}
\centering
\caption{Input parameters for SED fitting with CIGALE.}
\begin{tabular}{@{}lccc@{}}
\hline
\hline
Parameter&Symbol& Values\\
\hline
Metallicity   & Z& 0.02 (solar metallicity)\\
Age in Gyr & $t$& 0.01 0.05 0.1 0.5 1.0 2.0\\
e-folding time $\tau$ in Gyr & $\tau$& 0.1 0.5 1.0 5.0 10.0\\
\hline
V-band attenuation  & $A_{V}$ &0.5 1.0 1.5 2.0 2.5 3.0 3.5 4.0 4.5 5.0 5.5 6.0\\
Slope correction of the Calzetti attenuation law & $\delta$& -0.2 -0.1 0.0 0.1 0.2 0.5\\
\hline
IR power-law slope  of dust mass distribution over heating intensity & $\alpha$ & 1.2; 1.4; 1.5; 1.6; 1.7; 1.9\\
\hline
\end{tabular}
\label{tab:SEDparameter}
\end{table*}

\section{Discussion: SFR\--M$_*$ relation  --- a comparison with star-forming main sequence }
\label{sec:discussion}

\subsection{SFR\--M$_*$ relation}

Galaxy surveys at low and high redshifts have shown that star-forming galaxies form a power-law relation between their star formation rate and stellar mass, known as the ``main sequence" (MS) of star-forming galaxies. The six SPT DSFGs presented in this paper all show intense star formation spanning a redshift range from $z=2.8$ to $z=5.7$. We compare our results (Fig. \ref{fig:SFRmstar}) with previous determinations of the  $SFR\--M_*$ relations of \cite{daddi07} at $z$ = 2 and \cite{magdis10}  at $z$ = 3 to investigate whether SPT DSFGs lie on top of the main-sequence or are extreme outliers.  At $z$ = 2, \cite{daddi07} derived the relation based on a sample of massive galaxies selected with $K < 22$ from GOODS and 24$\mu$m observations, which is widely used as the fiducial main sequence at $z\sim2$.  \cite{magdis10} based their study on LBGs with IRAC observations, and found an increasing SFR at a given mass from $z$ = 2 to $z$ = 3. All the stellar masses and SFRs are calibrated to a Chabrier IMF (similar to Kroupa) and derived using the BC03 models. We do not correct our SMGs for the small differences between BC03 and M05 in the figure. 

SPT0538-50, SPT0103-45, and SPT2134-50 at $z\sim3$ lie above the $z$ = 3 main sequence by factors of 3, 10, and at least 3, respectively.  In contrast, the extensively studied \cite{chapman05} SMGs at $z$ = 2-3, even though known to have widely varying stellar masses depending upon different assumed star formation histories, suggest that most SMGs represent the high-mass extension of the main-sequence of star-forming galaxies at $z$ $\geq$ 2 (\citealt{michalowski12}). Here we show the results assuming the M05 models for direct comparison with our SMGs. Comparing the median values of the two samples, SPT SMGs have a factor of $\sim$ 2 higher SFRs and a factor of $\sim$ 2 lower stellar masses than those of the Chapman05 SMGs. This discrepancy still holds in the case of a double-SFH model (Fig. \ref{fig:SFRmstar}). This is suggestive of heterogeneity between SMG populations or two star formation modes: more quiescently star-forming SMGs and starburst SMGs (more discussion later in this section). We also show the individual SMGs from the ALESS sample \citep{dacunha15} which is not a uniform population: a significant fraction of the SMGs lie above the MS while some are consistent with being at the high-mass end of the MS.

\subsection{Specific SFRs}
\label{sec:ssfr}

The specific SFR ($SFR/M_*$), which can be interpreted as the inverse of the characteristic timescale for star formation to build the galaxy, is now recognized as a key parameter linked with the evolutionary status of these galaxies. Another way to explore the $SFR\--M_*$ relation is to look at sSFR as a function of redshift.  The sSFRs for the six sources are presented along with stellar mass and SFR estimates in Table \ref{tab:stellarmass}. We note that sSFRs are independent of lensing magnifications as long as differential magnification is not a concern.

\begin{figure*}
\centering
{\includegraphics[width=13cm, height=9cm]{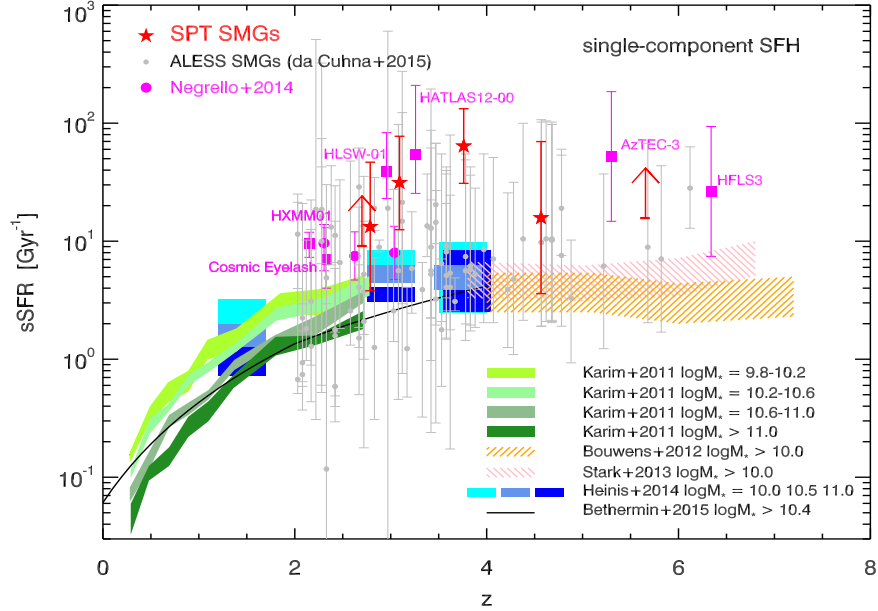}}    
\caption{sSFR -- $z$ relation. The red filled stars are the SPT DSFGs. SPT2134-50 and SPT0346-52 have lower limits on sSFRs (SPT2134-50 is shifted by -0.08 in redshift for clarity). We show the results from \cite{karim11} (green bands) which cover a board range of redshift and stellar mass bins. The blue shaded regions are average sSFRs obtained by \cite{Heinis14}  at $z$ $\sim$ 1.5, $z$ $\sim$ 3, and $z$ $\sim$ 4. The pink and orange shaded regions are the extrapolated relations by \cite{bouwens12} and \cite{stark13}. The black curve is a fitting formula of the main-sequence sample of star-forming galaxies up to $z$ = 4 in the COSMOS field by \cite{Bethermin15a}. The magenta squares are individual lensed DSFGs from the literature with sSFRs high above the MS: Cosmic Eyelash (i.e., SMMJ2135-0102) at $z$ = 2.3 \citep{Swinbank10}, HXMM01 at $z$ = 2.3 (shifted by -0.15 in $z$ for clarity; \citealt{fu13}), HLSW-01 at $z$ = 2.96 \citep{Conley11}, HATLAS12-00 at $z$ = 3.2 \citep{Fu12}, AzTEC-3 at $z$ = 5.3 \citep{capak11}, and HFLS3 at $z$ = 6.34 \citep{Cooray14}. The magenta circles are SDP.17, SDP.81, and SDP.130 from {\it Herschel}-ATLAS \citep{Negrello14}. The grey dots are $z$ $>$ 2 SMGs from the ALESS sample \citep{dacunha15}. SPT0103-45 and SPT2147-50 have significantly higher sSFRs, more than a factor of 10 above the MS, satisfying the criterion for strong starbursts of $\geq$ 10 sSFR$_{MS}$ defined in \cite{Bethermin15a}. }
\label{fig:sSFRmore}
\end{figure*}

We compare our results with literature on the $sSFR\--z$ relations (Fig. \ref{fig:sSFRmore}; see \citealt{Heinis14} and references therein). We show the results of \cite{karim11}, which span a wide stellar mass range at 0 $<$ $z$ $<$ 3,  and measurements from \cite{Heinis14}, which extend from $z$ = 1.5 to 4.  These studies establish an upper boundary on the sSFR-$z$ relations of normal star-forming galaxies in the literature. The dashed line is a fitting formula of the main-sequence sample of star-forming galaxies up to $z$ = 4 in the COSMOS field by \cite{Bethermin15b}. The relations from these studies enable us to make a fair comparison given that we make the same assumptions for IMF, SFH, and dust attenuation law in deriving the stellar masses, except that they adopt the BC03 model while we use M05. We stress that this has negligible effect on our conclusions. 

Following \cite{Heinis14}, we demonstrate the extrapolation of the results of \cite{bouwens12} and \cite{stark13} from $z$ = 4 to 7 at the relevant stellar mass bins by assuming that a power-law relation holds between SFR and $M_*$ at $z$ $>$ 4 but with a variation in the slope between 0.7 and 1. We have seen the anti-correlation between sSFR and $M_*$ for the star-forming samples. SPT0103-45 and SPT2147-50 have significantly higher sSFRs, more than a factor of 10 above the MS, satisfying the criterion for strong starbursts of $\geq$ 10 sSFR$_{MS}$ defined in \cite{Bethermin15b}.  They significantly deviate from the main-sequence star formation mode. SPT0538-50 lies about a factor of 4 above the MS and SPT2146-55 has a sSFR about 4 times higher with the lower limit reaching the MS. For SPT0346-52 and SPT2134-50, we only have lower limits that already exclude MS SFRs. The mean sSFR for the four detections lies above the MS at $\sim$ 5$\sigma$ confidence level. SPT SMGs, on average, have sSFRs $\sim$ 4 times higher than those of ALESS SMGs \citep{dacunha15} although a significant fraction of them are high-sSFR outliers. These galaxies are not a uniform population.  

Hydrodynamical simulations with dust radiative transfer calculations (e.g., \citealt{hayward12})  have suggested the `bimodality' of the SMG population: a mix of quiescently star-forming and starburst galaxies.  The quiescent SMGs are star-forming disks blended into one submillimeter source at early stages of a merger whereas at late stages (near coalescence) tidal torques drive gas inflows and trigger strong starbursts. We are likely witnessing the ongoing strong starburst events. 

So far we have been applying the same lensing magnification factors derived from ALMA (i.e., dust emission) to stellar emission. However, the effect of differential lensing between different wavelength regimes  can be prominent in galaxy-galaxy lensing due to spatial variations within the background galaxy.  Simulations have predicted this effect in galaxy-galaxy SMG systems (\citealt{hezaveh12a,serjeant12}) but we do not have the capability to measure it for our sample. \cite{Calanog14}, which presents the NIR-derived lens models of {\it Herschel}-selected galaxies, tests for differential lensing between the stellar and dust components. They find that the dust magnification factor measured at 880 $\micron$ is $\sim$ 1.5 times on average higher than the near-IR magnification factor. \cite{Dye15} and \cite{Rybak15} find a similar differential magnification factor, based on the analysis of the ALMA high resolution imaging of SDP.81.  Assuming our SPT DSFGs have similar differential lensing effect, the resultant sSFRs would decrease by $\sim$1/3 which are within the error bars.

Models predict that the SPT selection should identify a diverse set of sources. However, \cite{Bethermin15a} predicts that 90\% of SPT lensed sources are classified as main sequence galaxies. Even though the sample used for this study is small, the presence of extreme starbursts is thus surprising. One contributing factor may be the magnifications of the sources included in this subsample. Four of our sources have $\mu < 7$, while the average $\mu$ of the full SPT sample predicted by \cite{bethermin12b} is $\sim$15.  \cite{bethermin12b} predict that sources with $4 < \mu < 7$ on average lie a factor of 4.5 above the main sequence, compared to a factor of 1.7 for sources with $\mu > 20$.

\begin{table*}
\centering
\caption{Derived properties from the CIGALE SED fitting}
\begin{tabular}{@{}lcccccc@{}}
\hline
\hline
SPT source name    & stellar mass & L$_{IR}$ [8 -- 1000 $\mu$m]  & age  & $SFR_{CIGALE}$ & $SFR_{IR}$ & sSFR \\
ID            		        &   $M_{\Sun}$   & $L_{\Sun}$    &   Myr  &${\rm M_{\Sun}/yr}$    & ${\rm M_{\Sun}/yr}$  &   ${\rm Gyr^{-1}}$                           \\
\hline
SPT0103-45             &  $5.5 ^{+ 6.1}_{- 2.9}\times10^{10}$    &   $(1.2\pm0.1)\times10^{13}$  & $40 ^{+ 60}_{- 20}$  &$1740 ^{+ 380}_{- 310}$ &  1290\---1800    &   $31.4 ^{+ 46.1}_{- 18.9}$     \\
SPT0346-52 		& $< 3.1\times10^{11}$                          &  $(3.6\pm0.3)\times10^{13}$  & $30 ^{+ 50}_{- 20}$   &$4840 ^{+ 1090}_{- 890}$   & 3830\---5340 &   $> 15.7$   \\
SPT0538-50  		& $3.9 ^{+ 8.0}_{- 2.6}\times10^{10}$    &  $(4.3\pm0.5)\times10^{12}$ & $80 ^{+ 250}_{- 60}$  &$510 ^{+ 190}_{- 140}$     & 460\---640        & $13.3 ^{+ 33.4}_{- 9.5} $	\\
SPT2134-50		& $< 6.6 \times10^{10}$                         &  $(4.9\pm0.7)\times10^{12}$  & $40 ^{+ 70}_{- 20}$   &$640 ^{+ 230}_{- 160}$   & 520\---730          &  $> 9.1$     \\
SPT2146-55             & $0.8 ^{+ 1.9}_{- 0.6}\times10^{11}$    &  $(9.3\pm1.2)\times10^{12}$  & $70 ^{+ 250}_{- 50}$  &$1190 ^{+ 450}_{- 320}$   & 990\---1380      & $15.8 ^{+ 54.0}_{- 12.2}$    \\
SPT2147-50            & $2.0 ^{+ 1.8}_{- 0.9}\times10^{10}$    &  $(8.4\pm1.0)\times10^{12}$  & $20 ^{+ 30}_{- 10}$     &$1290 ^{+ 320}_{- 250}$   & 900\---1250      & $64.0 ^{+ 68.6}_{- 33.1}$     \\
\hline
\end{tabular}
\tablecomments{The derived properties are corrected for lensing magnifications. For SPT2134-50 and SPT0346-52 which are non-detections in the IRAC bands, we place 3$\sigma$ upper limits on the stellar masses. We note that specific SFRs are independent of magnification factors.}
\label{tab:stellarmass}
\end{table*}

\section{Summary and Conclusions}
\label{sec:conclusions}

We have determined stellar masses and star formation rates for six strongly lensed DSFGs using ALMA, {\it HST}, and {\it Spitzer} follow-up observations. The sources are drawn from the SPT DSFG program. 

We develop a de-blending technique to extract emission of the background DSFG when blended with that from the foreground lensing galaxy in {\it Spitzer}/IRAC imaging. {\it HST} WFC3 images are used to model the light profiles of the lenses which are subtracted in IRAC to produce the residual map overlaid with ALMA submillimeter imaging, indicating the position and structure of the gravitationally lensed DSFG. We demonstrate that in a gravitational lensing system like SPT0538-50, where the background DSFG is strongly magnified and distorted by a massive lensing halo to well separate lensing arcs as seen by ALMA,  a full Einstein ring structure is recovered in the residual map in both IRAC bands.  Sufficient angular separation is critical in terms of successfully disentangling  the mixed emissions from the lens and source. This is best illustrated in SPT2134-50.  In this case, there exists flux coincident with the ALMA contours, but we are unable to exclude the possibility that this signal is simply residual emission associated with the foreground lens.  A minimum separation, comparable to the FWHM of the PSF, is required to conduct the de-blending and extract robust fluxes. 

Based on the multi-wavelength photometry,  we have performed SED fitting with CIGALE. This fitting includes stellar synthesis models as well as dust emission templates (i.e., taking the full spectral energy distribution into consideration) to derive stellar masses, total IR luminosities, and instantaneous star formation rates.  The best-fit model is based on a single extended burst SFH.  The derived stellar masses span a wide range with a median value of  $\sim$ 5 $\times$ 10$^{10}$M$_{\Sun}$. The instantaneous SFRs range from 510 to 4800 $M_{\Sun} {\rm yr}^{-1}$. Use of a two-component SFH generally leads to higher stellar masses, higher SFRs but similar sSFRs, and our qualitative results are robust to choice of SFHs. 

We investigated the $SFR\--M_{*}$ relation and evolution of specific SFR with redshift. Compared to sSFR -- $z$ relations of normal star-forming galaxies in the literature, the six SPT DSFGs all  lie above the MS.  Two of them are $\sim$ 10 times above  the star-forming main-sequence. SPT0346-52 converts the gas into stars at a rate of $\sim$ 4800 $M_{\Sun}{\rm yr}^{-1}$ only 1 $G{\rm yr}$ after the Big Bang, a SFR that is among the highest at any epoch.  Our results suggest that we may be witnessing strong starburst events. One possible explanation for such extreme starbursts is gas-rich major mergers (e.g., \citealt{tacconi06,narayanan09,engel10,Dye15})  in which tidal torque compresses gas reservoir leading to a concentrated starburst. High resolution kinematical and morphological observations are needed to test the merger scenario. 

Future work involves determining dynamical mass from CO line width and gas mass from CO luminosity together with dust mass estimates to provide a more comprehensive interpretation of this unique population (Spilker et al. and Strandet et al. in preparation). As noted by others (\citealt{michalowski12, michalowski14}), deriving stellar masses for young, dusty DSFGs is an exercise fraught with challenge. For high redshift sources, the lack of rest-frame NIR bands further limits the achievable fidelity. Observations with JWST offer the potential for dramatic improvement due to the two-fold advantage of improved spatial resolution and greater wavelength coverage extending into the mid-infrared. In the MIR, which probes rest-frame Ks for distant SMGs, the foreground lens fades while the SMG brightens.  The improved spatial resolution compared to {\it Spitzer} greatly simplifies separation of source and lens fluxes, even for compact lensing configuration. Moreover, the required integration times will be short. For instance all the sources are expected to be detected within 1 minute integration time at S/N $\ge$ 10 at rest-frame Ks band.  \newline\newline

\section*{acknowledgments}

We thank the anonymous referee for useful comments. We thank Sebastien Heinis for sharing the data in \cite{Heinis14}. 
This work is based on observations made with the NASA/ESA Hubble Space Telescope, obtained at the Space Telescope Science Institute, which is operated by the Association of Universities for Research in Astronomy, Inc., under NASA contract NAS 5-26555. These observations are associated with programs \#12659 and \#13614.
Support for programs \#12659 and \#13614 was provided by NASA through a grant from the Space Telescope Science Institute, which is operated by the Association of Universities for Research in Astronomy, Inc., under NASA contract NAS 5-26555.
This work is based in part on observations made with the Spitzer Space Telescope, which is operated by the Jet Propulsion Laboratory, California Institute of Technology under a contract with NASA. Support for this work was provided by NASA through an award issued by JPL/Caltech.
Partial support for this work was provided by NASA from awards for Herschel observations for OT1\_dmarrone\_1, OT1\_jvieira\_4, and OT2\_jvieira\_5.
Herschel is an ESA space observatory with science instruments provided by European-led Principal Investigator consortia and with important participation from NASA.
J.S.S., D.P.M., and J.D.V. acknowledge support from the U.S. National Science Foundation under grant No. AST-1312950 and through award SOSPA1-006 from the NRAO.
M. A. acknowledges partial support from FONDECYT through grant 1140099. 
This paper makes use of the following ALMA data: ADS/JAO.ALMA $\#$2011.0.00957.S and $\#$2011.0.00958.S. ALMA is a partnership of ESO (representing its member states), NSF (USA) and NINS (Japan), together with NRC (Canada) and NSC and ASIAA (Taiwan), in cooperation with the Republic of Chile. The Joint ALMA Observatory is operated by ESO, AUI/NRAO and NAOJ. 
The SPT is supported by the National Science Foundation through grant ANT-0638937, with partial support through PHY-1125897, the Kavli Foundation and the Gordon and Betty Moore Foundation. 
This research has made use of NASA's Astrophysics Data System.

\bibliographystyle{apj}
\bibliography{spt_smg}

\end{document}